\documentclass{article} 
\usepackage{iclr2026_conference,times}

\usepackage{xspace}
\usepackage{mathtools}
\usepackage{tikz}
\usepackage{multirow}
\usepackage{multicol}
\usepackage[dvipsnames, table]{xcolor}
\usepackage{subfig}
\usepackage{booktabs}
\usepackage{hyperref}
\usepackage{url}
\usepackage[ruled,lined]{algorithm2e}
\usepackage{fancyhdr}

\graphicspath{{figs/}}

\author{%
Zheng Liu$^{1*}$ \quad He Zhu$^{1*}$ Xingyang Li$^1$  \quad Yirun Wang$^1$ \quad Yujiao Shi$^3$ \quad Yiming Gan$^4$ \\
\textbf{Wei Li}$^1$ \quad \textbf{Jingwen Leng}$^{1,2}$ \quad \textbf{Yu Feng}$^{1,2,\#}$ \quad \textbf{Minyi Guo}$^{1,2}$\\
$^1$Shanghai Jiao Tong University \quad $^2$Shanghai Qizhi Institute  \\
$^3$ShanghaiTech University \quad $^4$Chinese Academy of Sciences
\\
$^*$Equal Contribution \quad $^\#$Corresponding Author\\
\texttt{y-feng@sjtu.edu.cn}\\
Project site: \texttt{\url{https://voyager-web.netlify.app/}}
}

\usepackage{amsmath,amsfonts,bm}

\def\eqref#1{equation~\ref{#1}}

\def\1{\bm{1}}

\DeclareMathAlphabet{\mathsfit}{\encodingdefault}{\sfdefault}{m}{sl}
\SetMathAlphabet{\mathsfit}{bold}{\encodingdefault}{\sfdefault}{bx}{n}

\def\gF{{\mathcal{F}}}

\def\gS{{\mathcal{S}}}

\def\sI{{\mathbb{I}}}

\def\sR{{\mathbb{R}}}
\def\sS{{\mathbb{S}}}

\newcommand{\goldtext}[1]{{\textcolor{YellowOrange}{\textbf{#1}}}\xspace}
\newcommand{\silvertext}[1]{{\textcolor{Gray}{\textbf{#1}}}\xspace}

\ifx\figurename\undefined \def\figurename{Figure}\fi
\renewcommand{\figurename}{Fig.}

\newcommand{\Sect}[1]{Sec.~\ref{#1}}
\newcommand{\Fig}[1]{Fig.~\ref{#1}}
\newcommand{\Tbl}[1]{Tbl.~\ref{#1}}

\newcommand{\Alg}[1]{Algo.~\ref{#1}}

\newcommand{\specialcell}[2][c]{\begin{tabular}[#1]{@{}c@{}}#2\end{tabular}}

\renewcommand{\subparagraph}[1]{\underline{\textit{#1}}\xspace}

\newcommand{\proj}{\textsc{Voyager}\xspace}
\newcommand{\mode}[1]{\underline{#1}\xspace}

\newcommand{\RNum}[1]{\uppercase\expandafter{\romannumeral #1\relax}}

\newcounter{alphasect}
\def\alphainsection{0}

\let\oldsection=\section
\def\section{%
  \ifnum\alphainsection=1%
    \addtocounter{alphasect}{1}
  \fi%
\oldsection}%

\renewcommand\thesection{%
  \ifnum\alphainsection=1%
    \Alph{alphasect}
  \else%
    \arabic{section}
  \fi%
}%

\newenvironment{alphasection}{%
  \ifnum\alphainsection=1%
    \errhelp={Let other blocks end at the beginning of the next block.}
    \errmessage{Nested Alpha section not allowed}
  \fi%
  \setcounter{alphasect}{0}
  \def\alphainsection{1}
}{%
  \setcounter{alphasect}{0}
  \def\alphainsection{0}
}%

\title{\proj: Real-Time City-Scale 3D Gaussian Splatting on Resource-Constrained Devices}

\author{%
Zheng Liu$^{1*}$ \quad He Zhu$^{1*}$ Xingyang Li$^1$  \quad Yirun Wang$^1$ \quad Yujiao Shi$^3$ \quad Yiming Gan$^4$ \\
\textbf{Wei Li}$^1$ \quad \textbf{Jingwen Leng}$^{1,2}$ \quad \textbf{Yu Feng}$^{1,2,\#}$ \quad \textbf{Minyi Guo}$^{1,2}$\\
$^1$Shanghai Jiao Tong University \quad $^2$Shanghai Qizhi Institute  \\
$^3$ShanghaiTech University \quad $^4$Chinese Academy of Sciences
\\
$^*$Equal Contribution \quad $^\#$Corresponding Author\\
\texttt{y-feng@sjtu.edu.cn}\\
Project site: \texttt{\url{https://voyager-web.netlify.app/}}
}

\iclrfinalcopy

\begin{document}

\maketitle

\begin{abstract}

3D Gaussian splatting (3DGS) is an emerging technique for photorealistic 3D scene rendering. 
However, rendering city-scale 3DGS scenes on resource-constrained mobile devices in real-time remains a significant challenge due to two compute-intensive stages: level-of-detail (LoD) search and rasterization.

In this paper, we propose \proj, an effective solution to accelerate city-scale 3DGS rendering on mobile devices. 
Our key insight is that, \textit{under normal user motion, the number of newly visible Gaussians within the view frustum remains roughly constant}. 
Leveraging this temporal correlation, we propose a temporal-aware LoD search to identify the necessary Gaussians for the remaining rendering stages. 
For the remaining rendering process, we accelerate the bottleneck stage, rasterization, via preemptive $\alpha$-filtering.
With all optimizations above, our system can deliver low-latency, city-scale 3DGS rendering on mobile devices.
Compared to existing solutions, \proj achieves up to 6.6$\times$ speedup and 85\% energy savings with superior rendering quality.
Codes will be released upon publication.

\end{abstract}

\section{Introduction}
\label{sec:intro}

The field of 3D Gaussian splatting (3DGS)~\citep{kerbl20233d, fan2023lightgaussian, fang2024mini, girish2024eagles, lee2024compact, niemeyer2024radsplat, wang2024adr, mallick2024taming, gui2024balanced, radl2024stopthepop} has made tremendous progress in recreating the 3D world with photo-realistic fidelity. 
While 3DGS unlocks new possibilities and fuels our imagination, rendering city-scale 3DGS scenes on mobile devices, e.g., the smartphone in your pocket, remains a substantial challenge. 
Primarily, there are some inherent limitations of mobile devices, such as limited compute resources and memory bandwidth.
For example, a single scene like \textit{SmallCity} from HierarchicalGS~\citep{kerbl2024hierarchical} can barely achieve 12.2 frame-per-second (FPS) on a Nvidia mobile Ampere GPU~\citep{orinsoc}.
These constraints make it infeasible to achieve city-scale 3DGS on today's mobile devices.

In order to render massive-scale scenes, 3DGS algorithms require managing tens of millions of Gaussian ellipsoids, a.k.a. Gaussians. 
For instance, \textit{SmallCity} from HierarchicalGS~\citep{kerbl2024hierarchical} contains over $ 1.8 \times 10^7 $ Gaussians.
To effectively manage at this scale, the state-of-the-art city-scale 3DGS methods~\citep{kerbl2024hierarchical, ren2024octree, liu2024citygaussian} often adopt a two-stage pipeline: \textit{level-of-detail (LoD) search} and \textit{splatting}. 
LoD search uses a hierarchical representation to manage those Gaussians and selects a subset of Gaussians at an appropriate LoD under the current camera pose.
Splatting then renders these selected Gaussians on the canvas in a visually consistent order.
Our experiment in \Fig{fig:exec_time} shows that both stages take non-trivial execution time, with the LoD search further dominating as the scene complexity increases.

To reduce the computation workload of LoD search, our key insight is that, as the user navigates through a 3D scene, \textit{the number of newly visible Gaussians introduced by continuous viewpoint changes remains roughly constant} (assuming a general case of a smooth user motion). 
The subsets of selected Gaussians across adjacent frames are often similar in LoD search.
By leveraging this insight, we propose a \textit{temporal-aware LoD search} algorithm in \Sect{sec:method:search}.
Specifically, we propose two GPU-oriented optimizations in LoD search.
For the initial frame, we propose an efficient tree traversal algorithm that can be highly parallelized on GPUs to accelerate LoD search.
For the subsequent frames, we design a companion data structure that leverages temporal correlations across consecutive tree traversals to further accelerate this process.
Rather than recomputing the tree traversal results from scratch, our data structure can reuse previous traversal results to reduce the computation.
Compared against the previous LoD search algorithms~\citep{kerbl2024hierarchical}, we achieve 6.5$\times$ speedup with the bit-accurate LoD search result.

To improve the efficiency of splatting, prior studies~\citep{wang2024adr, fan2023lightgaussian, fang2024mini, huang2025seele, lin2025metasapiens} have primarily focused on reducing the total number of Gaussians in rasterization, a sub-step in splatting.
However, these methods often overlook the underlying bottleneck within rasterization itself.
We pinpoint the major bottleneck in rasterization and propose a GPU-oriented optimization, \textit{preemptive $\alpha$-filtering} in \Sect{sec:method:raster}, to accelerate rasterization.
Our key observation is that the primary source of slowdown during rasterization stems from the per-ray computation of Gaussian transmittance, $\alpha_i$. 
This operation is computationally expensive due to the heterogeneous design of modern GPU architectures (see \Sect{sec:motiv:perf}).
To address this issue, our filter can pre-determine the subset of Gaussians that do not need to compute transmittance, $\alpha_i$, without explicitly calculating $\alpha_i$.
For those Gaussians that do require $\alpha$-computation, we bypass the costly $\alpha$-computation by embedding a 128-byte precomputed lookup table (LUT) in the GPU's shared memory and approximating Gaussian transmittance instead.

Together, our algorithm can achieve up to 6.6$\times$ speedup compared to prior city-scale 3DGS algorithms on a mobile platform, Nvidia Orin SoC.
The contributions of this paper are as follows:
\begin{itemize}
    \item A parallel LoD search algorithm that exploits temporal coherence across frames to avoid repetitive tree traversal in LoD search.
    \item A GPU-efficient rasterization method that bypasses the key bottleneck in rasterization and significantly improves rendering efficiency.
    \item A framework that enables rendering city-scale 3DGS models on mobile devices, achieving up to 6.6$\times$ speedup and 85\% energy savings compared to prior works.
\end{itemize}

\section{Related Work}
\label{sec:related}

\paragraph{Efficient 3DGS.}
State-of-the-art methods for the efficiency of 3DGS can largely be categorized into two main directions.
The first direction focuses on reducing model size. 
Several studies propose various pruning techniques~\citep{fan2023lightgaussian, fang2024mini, girish2024eagles, niemeyer2024radsplat, feng2024evsplitting, peng2024rtg, tu2024fast, franke2024vr, hahlbohm2025efficient}, while others introduce alternative representations to minimize storage requirements~\citep{roessle2024l3dg, lee2024compact, huang20242d, wang2024simple, dai2024high, lee2024compact, girish2024eagles}.

The second direction aims to improve the efficiency of the 3DGS rendering pipeline. 
Some approaches introduce fine-grained filtering techniques for Gaussian-ray intersections to alleviate the computation of subsequent stages~\citep{feng2024flashgs, lee2024gscore, wang2024adr}. 
For instance, various online filtering techniques, such as axis-aligned bounding box (AABB)~\citep{wang2024adr} and oriented bounding box (OBB)~\citep{lee2024gscore} intersection tests, have been proposed for more fine-grained filtering.
Others tackle GPU performance bottlenecks such as warp divergence and workload imbalance in rasterization~\citep{huang2025seele, gui2024balanced, mallick2024taming}.

In contrast to these prior works, our framework focuses on large-scale 3DGS rendering. We jointly optimize both the LoD search and splatting stages, while addressing the performance constraints and hardware limitations that exist in today’s mobile GPUs.

\paragraph{Large-scale 3DGS.}
How to represent Level-of-Detail (LoD) efficiently is an issue existed broadly in neural rendering~\citep{turki2023pynerf, xu2023vr, barron2023zip}. 
Since 3DGS models are inherently point-based representations, various tree-based structures have been designed to manage large-scale scenes, such as Octree-GS~\citep{ren2024octree}, CityGaussian~\citep{liu2024citygaussian}, and HierarchicalGS~\citep{kerbl2024hierarchical}. 
These hierarchical structures organize Gaussian points to support scalable rendering. 
However, the irregular tree traversals introduce non-trivial runtime overhead.

In contrast, our LoD search is co-designed with GPU execution characteristics, such as the SIMT execution model.
With the temporal correlation that commonly exists in continuous rendering, our method can achieve more efficient tree traversal and improve the overall runtime performance.

\begin{figure*}[t]
    \centering
    \includegraphics[width=0.95\textwidth]{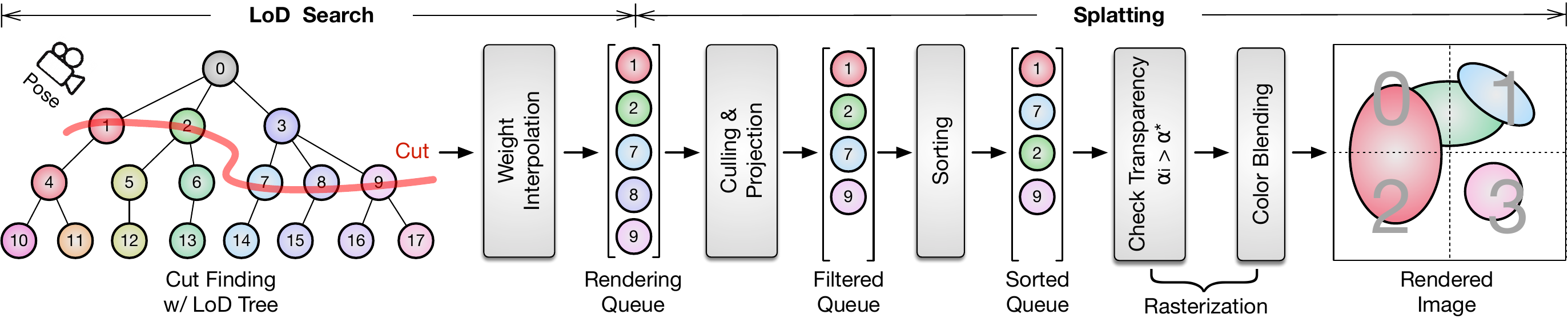}
    \caption{Overview of rendering pipeline for large-scale 3D Gaussians. LoD search consists of two sub-steps: cut finding and weight interpolation. Splatting contains three sub-steps: projection, sorting, and rasterization. 
    LoD search traverses the LoD tree to determine a set of Gaussians, given a LoD granularity.
    The result Gaussians form a ``cut'' that separates the top and bottom of the LOD tree.
    Then, the Gaussians on the cut go through a sequence of operations to render an image.
}
    \label{fig:hgs_pipeline}
\end{figure*}

\section{Motivation}
\label{sec:motiv}

In this section, we first introduce the general pipeline of the large-scale 3DGS algorithm in \Sect{sec:motiv:pre}. 
We then analyze the performance bottlenecks of existing algorithms on modern GPUs in \Sect{sec:motiv:perf}.

\subsection{Preliminary}
\label{sec:motiv:pre}

\paragraph{LoD Tree.}  
All city-scale 3DGS algorithms require a hierarchical tree-like structure to represent the entire scene.
This data structure serves two main purposes: first, it identifies the Gaussians inside the view frustum; second, it supports rendering at an appropriate level of detail.
Here, we describe the most general form of this data structure, \textit{LoD tree}, where each tree level represents a specific detail granularity (see \Fig{fig:hgs_pipeline}). 
Each tree node contains a single Gaussian with an unfixed number of child nodes.
Gaussians in lower levels are generally smaller and provide finer details.

\paragraph{Pipeline.} 
\Fig{fig:hgs_pipeline} gives an overview of the pipeline, which consists of two main stages: \textit{LoD search} and \textit{splatting}. 
Each of these stages includes several sub-steps, which we describe as follows:
\begin{itemize}
    \item{\textit{Cut Finding}}: This step traverses the LoD tree from top to bottom. At each node, we assess if the projected dimension of the Gaussian is smaller than the predefined LoD ($\tau^*$), while the projected dimension of its parent node is larger. We then gather all the Gaussians that meet this criterion. Essentially, these Gaussians form a ``cut'' that separates the top and bottom of the LoD tree, as shown in \Fig{fig:hgs_pipeline}.
    \item{\textit{Weight Interpolation}}: Each selected Gaussian then interpolates with its parent node to ensure a smooth transition across different LoDs. The interpolated Gaussians are inserted into the rendering queue in \Fig{fig:hgs_pipeline}.
    \item{\textit{Culling \& Projection}}: Once the cut is determined, the selected Gaussians are projected onto the image plane. Gaussians outside the view frustum or deemed irrelevant are culled.
    \item{\textit{Sorting}}: The remaining Gaussians are then sorted by depth, from the nearest to the farthest. 
    \item{\textit{Rasterization}}: The final step contains two parts. 
    First, each pixel calculates the intersected transmittance $\alpha_i$ for each Gaussian. If $\alpha_i$ is below a predefined threshold, $\alpha^*$, the Gaussian is skipped for color blending. Otherwise, the Gaussian contributes to the final pixel color via weighted blending.
\end{itemize}

\subsection{Performance Analysis}
\label{sec:motiv:perf}

\begin{figure}[t]
\centering
\begin{minipage}[t]{0.25\columnwidth}
  \centering
  \includegraphics[width=\columnwidth]{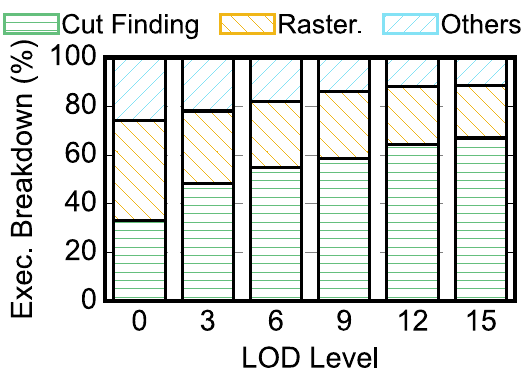}
  \caption{Normalized execution breakdown across different LoDs.}
  \label{fig:exec_time}
\end{minipage}
\hspace{1pt}
\begin{minipage}[t]{0.25\columnwidth}
  \centering
  \includegraphics[width=\columnwidth]{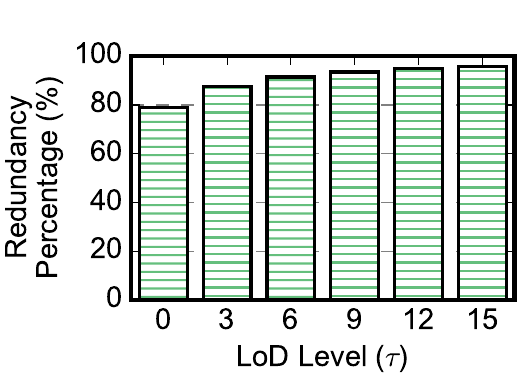}
  \caption{The percentage of redundant tree node accesses across LoDs.}
  \label{fig:redundant_pct}
\end{minipage}
\hspace{1pt}
\begin{minipage}[t]{0.20\columnwidth}
  \centering
  \includegraphics[width=\columnwidth]{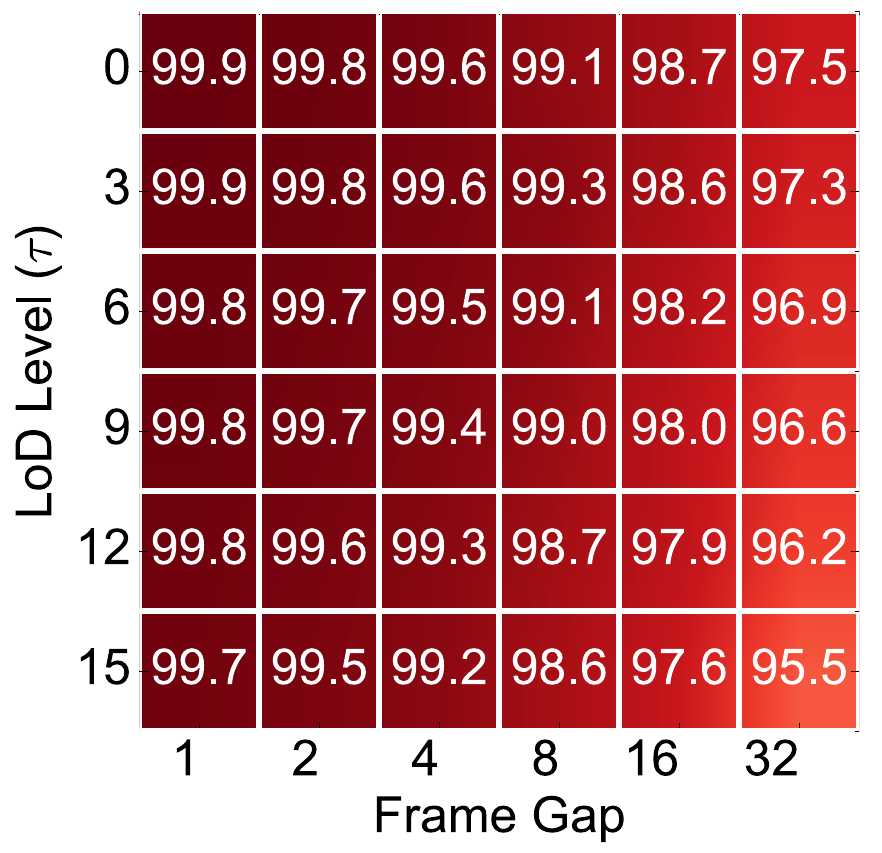}
  \caption{The overlap of selected Gaussians in the cuts between frames.}
  \label{fig:repetitive_comp}
\end{minipage}
\hspace{1pt}
\begin{minipage}[t]{0.25\columnwidth}
  \centering
  \includegraphics[width=\columnwidth]{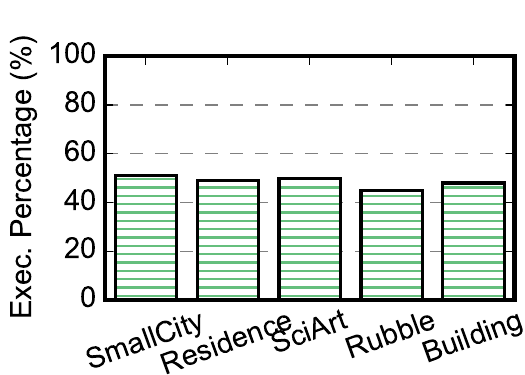}
  \caption{The execution percentage of $\alpha$ computation in rasterization.}
  \label{fig:exp_pct}
\end{minipage}
\end{figure}

\paragraph{Execution Breakdown.}
\Fig{fig:exec_time} shows the execution breakdown of HierarchicalGS across various LoDs on a Nvidia mobile Ampere GPU~\citep{orinsoc}.
Here, we use the \textit{SmallCity} scene from the HierarchicalGS dataset as an example.
When the LoD ($\tau^*$) is low, execution time is dominated by the rasterization step in the splatting stage. 
As the LoD increases, the cut finding step in the LoD search stage becomes the main contributor to overall execution time, accounting for up to 67\%. 
On average, cut finding and rasterization accounts for 83\% of total execution time. 
This means that both steps must be accelerated to achieve substantial speedups by Amdahl’s Law~\citep{amdahl1967validity}.

\paragraph{Cut Finding.}
We identify two primary inefficiencies in cut finding.
First, existing methods primarily rely on exhaustive search to find the cut, as the inherent irregularity of the LoD tree makes it difficult to achieve efficient parallelism on modern GPUs. 
Such a brute-force approach often leads to substantial redundant computation.
Our experiment shows that 92\% of tree node accesses are unnecessary when the LoD is 6 (see \Fig{fig:redundant_pct}).
Second, in real-time rendering scenarios, repeatedly performing tree traversal from the root node for every frame introduces considerable redundant computation. 
We simulate a 60 FPS real-time rendering scenario and analyze the overlap in selected Gaussians across consecutive frames. 
\Fig{fig:repetitive_comp} shows that 99\% of the selected Gaussians remain unchanged across frames when the LoD is 6. 
This indicates that a substantial amount of computation is wasted due to repetitive tree traversal.

\paragraph{Rasterization.}
While a few studies have addressed certain inefficiencies in rasterization, such as taming workload imbalance~\citep{wang2024adr} and reducing warp divergence~\citep{huang2025seele}, existing works have largely overlooked the most time-consuming operation in rasterization: computing the Gaussian transmittance, $\alpha_i$, for each pixel-Gaussian pair. 
Using Nvidia Nsight Compute~\citep{nsight}, \Fig{fig:exp_pct} shows that this operation takes roughly 50\% of the total rasterization time.
The root cause is the imbalanced compute units within modern GPUs. 
Taking Nvidia GPUs as an example, most operations can be massively parallelized by CUDA cores. 
However, the computation of Gaussian transmittance requires exponential computation, which is executed by special functional units (SFUs). 
Since CUDA cores vastly outnumber the SFUs (16:1)~\citep{ampere_arch}, calculating Gaussian transmittance becomes a key bottleneck.

\section{Methodology}
\label{sec:method}

To address the challenges in \Sect{sec:motiv:perf}, we present our framework, \proj, which introduces two key optimizations, temporal-aware LoD search (\Sect{sec:method:search}) and preemptive $\alpha$-filtering (\Sect{sec:method:raster}), targeting cut finding and rasterization in city-scale 3DGS rendering, respectively.

\subsection{Temporal-Aware LoD Search}
\label{sec:method:search}




We first explain our data structure, which is augmented on top of the original LoD tree to enhance GPU parallelism.
Next, we explain our fully-streaming LoD tree traversal that can streamingly process subtrees in parallel on off-the-shelf GPUs.
Finally, we describe how our temporal-aware LoD search can leverage previous cut results to further accelerate the LoD search.

\paragraph{LoD Tree Partitioning.} 
Recall, from \Sect{sec:motiv:pre}, each tree node in the LoD tree contains a single Gaussian with a variable number of child nodes.
An important performance issue with this irregular data structure is the workload imbalance across different GPU warps\footnote{A GPU warp is a group of threads that execute the same instruction in lockstep.}.
To address this issue, we propose to partition the LoD tree into smaller chunks, such that each chunk has a similar workload.

In our partitioning algorithm, we first split the entire LoD tree into $N$ small subtrees from the top to the bottom.
Each subtree is constrained to have a maximal tree height of 2.
Once the decomposition is complete, the resulting subtrees are sorted in a breadth-first search (BFS) order and stored in a list, $\sS = \{\gS_0, \gS_1, ..., \gS_N\}$, where each $\gS_i$ represents a single subtree.
With this transformation, the tree partitioning problem can be converted into a constrained bin packing problem~\citep{martello1990knapsack}.
The objective is to pack the subtrees into the minimum number of chunks, each corresponding to a batch of work for one GPU block.
The constraint is that each chunk must contain subtrees with contiguous IDs in $\sS$, to ensure the entire tree is traversed from the top to bottom.

\subparagraph{Optimization Objective.}
Given this setup, we can convert this bin packing problem into a linear constrained optimization.
The objective is to minimize the number of chunks:
\begin{equation}
    \underset{\Theta}{\operatorname{argmin}}\ \gF(\Theta) = \operatorname{Size}(\Theta),
    \ \text{where}\ \Theta = \{\theta_{i}\},\ \theta_{i} \in \{0, 1\},\ \text{and}\ i \in [0, 1, ..., N-1].
\label{eqn:obj}
\end{equation}
Here, each non-zero $\theta_i$ represents the starting index of a new chunk.
All non-zero $\theta_i$ are grouped into $\sI$.
Any two consecutive non-zero entries, $\theta_{n}$ and $\theta_{m}$, define a chunk that includes the substree IDs within $[n,m)$.
These subtrees are grouped and assigned to a single GPU block.

\subparagraph{Memory Constraints.}
To maximize usage of the on-chip memory on GPUs, we set the total size of each chunk to be smaller than the GPU shared memory size ($M$) per streaming multiprocessor,
\begin{align}
    \forall \theta_m \in \sI,\ & \sum_{\theta_i = \theta_m}^{\theta_{m+1}-1} \operatorname{SubtreeSize}(\theta_i)\leq M.
\label{eqn:local_dep}
\end{align}
This design ensures that all traversal operations for a chunk can be performed entirely within shared memory, avoiding off-chip memory stalls that would otherwise stall the GPU pipeline.
Given this formulation, we can easily apply existing integer linear programming solvers, such as Google OR-Tools~\citep{ortool}, to obtain the optimal LoD tree partitioning offline.

\begin{figure*}[t]
    \centering
    \includegraphics[width=\textwidth]{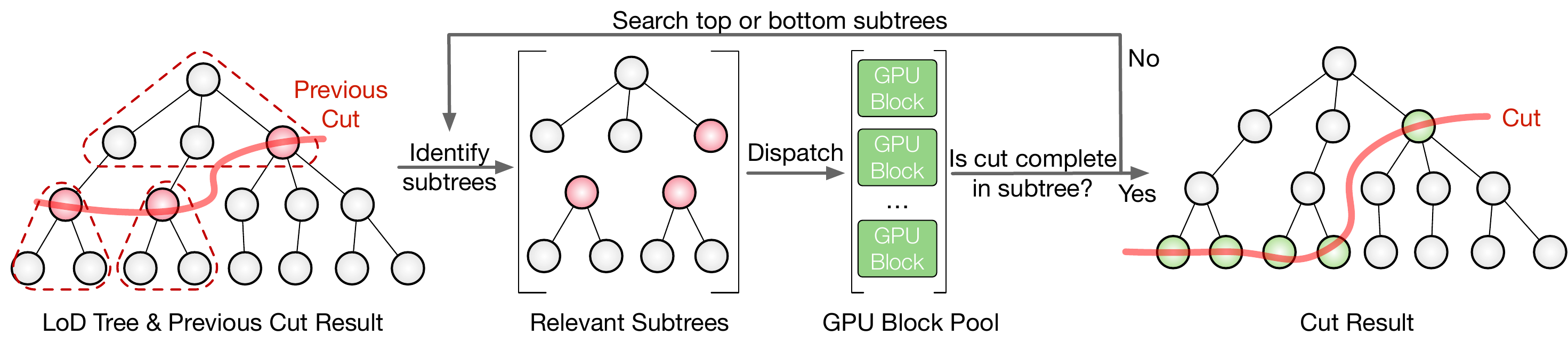}
    \caption{The overview of temporal-aware LoD search. Our algorithm exploits the temporal correlations across frames and reuses the previous ``cut'' result to avoid the redundant tree traversal.}
    \label{fig:temporal_aware_search}
\end{figure*}

\paragraph{Fully-Streaming LoD Tree Traversal.}
With our LoD tree partitioning, we propose a fully streaming algorithm that achieves high parallelism on GPUs while avoiding unnecessary tree node visits.
Rather than relying on the inherent parent-child relations in the original LoD tree, we perform tree traversal chunk by chunk in BFS order.
During the GPU execution, each chunk is assigned to a GPU warp. 
The tree nodes within one chunk are then evenly distributed among threads within a warp to ensure a balanced workload across GPU threads. 
Since each chunk of nodes is small enough to fit in GPU shared memory, our algorithm streamingly processes those tree nodes and avoids irregular global memory accesses in the original LoD tree traversal.

The workload assignment is dynamically dispatched whenever a warp becomes available at runtime. 
The tree traversal terminates once a clean cut separates the top and bottom of the LoD tree, thereby avoiding redundant tree accesses.
Meanwhile, we adopt the warp specialization~\citep{bauer2014singe} that uses a small group of dedicated GPU threads within each warp for data loading.
This allows data loading to be overlapped with computation, hiding the latency of loading LoD tree chunks.

\begin{algorithm}[h]
\caption{Algorithm of Temporal-Aware LoD Search}
\label{algo:subtree}
\KwData{LoD tree $ \sS $, Previous cut, $\sR_{prev}$, LoD granularity, $\tau^*$. }
\KwResult{cut $ \sR_{cut} $.}
$ \sS_{prev} \leftarrow \text{identifySubtreeInPrevCut}( \sS, \sR_{prev}, \tau^* )$\;
\While{$\sS_{prev}$ is not empty}{
  $\gS_i \gets \sS_{prev}.\text{dequeue()}$\;
  $ F_{signal}, \sR_{tmp} \gets \text{findCut(}\gS_i, \tau^* \text{)}$; \ // delegate to one GPU warp if available\;
  \For{ $n_i$ in $\sR_{tmp}$ }{
    $ \sR_{cut}\text{.enqueue(} n_i \text{)} $; \ // enqueue the subtree result\;
  }
  \If{$F_{signal}$ == \texttt{TOP} and $\sS_{prev}$\text{.parent}\ is\ not\ visited}{
    $\sS_{prev}\text{.push(} \sS_{prev}\text{.parent} \text{)}$; \ // too fine-grained, need to find its parent\; 
  }
  \If{$F_{signal}$ == \texttt{BOTTOM}}{
    \For{ $n_i$ in $\sR_{tmp}$ }{
      $\sS_{prev}\text{.push(}  \text{)}$; \ // too coarse-grained, need to find its children\; 
    }
  }
}

\end{algorithm}

\paragraph{Temporal-Aware LoD Search.} For tree traversal of subsequent frames, we further exploit the temporal correlations across frames and introduce a temporal-aware LoD search.
Our algorithm also leverages our LoD tree partitioning and works on the subtree traversal.
\Fig{fig:temporal_aware_search} highlights some individual subtrees in red dashed blocks.
Here, we only show a two-level subtree partitioning for illustration purposes; our actual implementation applies multi-level partitioning to the LoD tree.

The overall algorithm is also described in \Alg{algo:subtree}.
Given the cut result, $ \sR_{prev} $, from the previous frame (highlighted in pink in \Fig{fig:temporal_aware_search}), our algorithm first identifies the subtrees, $ \sS_{prev} $, to which each Gaussian in the cut belongs.
Then, our algorithm only traverses those identified subtrees instead of all subtrees in a GPU streaming fashion.
In GPU implementation, each subtree is assigned to a separate GPU warp for local subtree traversal.
Because each subtree is approximately equal in size, it ensures a balanced workload distribution across GPU warps.
If searching the local subtree cannot obtain the complete ``cut'' result, i.e., no clean cut for this subtree (signaled by $ F_{signal} $), we then search its corresponding parent subtree or child subtrees to complete the cut finding. 
This way, we ensure that the results from our temporal-aware LoD search are bit-accurate.

\subsection{Preemptive $\alpha$-Filtering}
\label{sec:method:raster}

In \Sect{sec:motiv:perf}, we show that the primary performance bottleneck in rasterization is the computation of Gaussian transmittance, which accounts for over 50\% of the total execution time. 
This operation introduces significant latency because it relies on exponential computations.
Unlike common arithmetic operations, exponentials are not executed by the massive CUDA cores.
Instead, they are executed by SFUs, which are far fewer in number compared to CUDA cores (1:16).
This imbalance in hardware resources makes transmittance computation a major bottleneck in rasterization.

To address this, we propose \textit{preemptive $\alpha$-filtering} to accelerate rasterization.
In canonical rasterization, the Gaussian transmittance, $\alpha$, is computed by,
\begin{equation}
    \alpha = \min(0.99, \theta_i \cdot e^\rho),
\end{equation}
where $\rho$ is the power decay of the current Gaussian-ray intersection and $\theta_i$ is the constant opacity of the intersected Gaussian $i$.
Each $\alpha_i$  is then compared against a transmittance threshold $\alpha^*$ to determine whether to proceed with the subsequent color blending.

Instead of computing the expensive exponential for every Gaussian-ray intersection, we refactor the transmittance comparison as,
\begin{equation}
\log(\alpha^*) - \log(\theta_i) < \rho.
\end{equation}
Since $\log(\theta_i)$ can be precomputed offline and stored for each Gaussian, we replace the original transmittance check with our lightweight check, $\log(\alpha^*) - \log(\theta_i) < \rho$, before computing the transmittance, $\theta_i \cdot e^\rho$. 
This optimization has two advantages.
First, it allows us to skip unnecessary exponent computations for Gaussians that are below the transmittance threshold, $\alpha^*$.
Second, by moving this preemptive check to the \textit{culling\&projection} stage, we can further reduce the computation overheads for subsequent stages: sorting and rasterization. 

Although our lightweight check reduces the number of exponential operations, it cannot eliminate them completely. 
To remove these computations altogether, we introduce a lookup table (LUT) embedded in GPU shared memory.
The LUT can directly map $\rho$ values to the corresponding exponential results, $e^\rho$, allowing us to obtain $e^\rho$ via a memory lookup rather than invoking SFUs.

\begin{figure}[t]
    \centering
    \includegraphics[width=0.6\columnwidth]{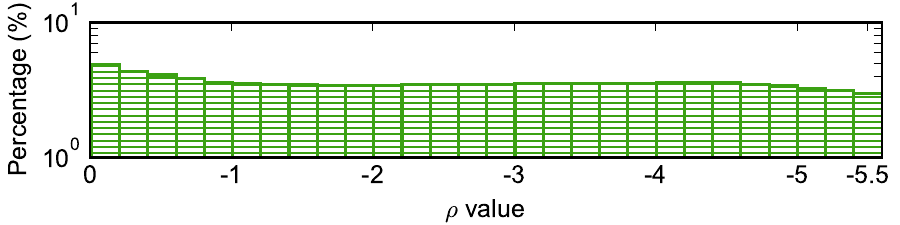}
    \caption{The distribution of $\rho$ values across the value space, $[-5.55, 0]$.}
    \label{fig:exp_dist}
\end{figure}

The next question is how to design such a LUT.
Our observation is that $\rho$ values are uniformly distributed between the effective value space, $[-5.55, 0]$, as shown in \Fig{fig:exp_dist}.
Thus, we uniformly split the effective value space into $m$ intervals. Here, we choose m to be 32.
a given $\rho$ value can be directly indexed to this LUT by computing,
\begin{equation}
    \text{index} = \left\lfloor \frac{\rho}{-5.55} \cdot m \right\rceil,
\end{equation}
where $\lfloor \cdot \rceil$ denotes rounding to the nearest integer.
The LUT indexing enables low-latency exponential approximation via a small shared memory.
The sensitivity of $m$ is shown in \Fig{fig:lut_size}.

\section{Evaluation}
\label{sec:eval}

\subsection{Experimental Setup}
\label{sec:eval:exp}

To show the efficiency and robustness of \proj, we evaluate on three large-scale datasets: HierarchicalGS~\citep{kerbl2024hierarchical}, UrbanScene3D~\citep{lin2022capturing}, and MegaNeRF~\citep{turki2022mega}.
Meanwhile, we also test \proj on small-scale datasets: Mip-NeRF360~\citep{barron2021mip}, Tank\&Temple~\citep{knapitsch2017}, and DeepBlending~\citep{hedman2018deep}.   
For quality evaluation, we use PSNR, SSIM, and LPIPS.
We also report frame-per-second (FPS), the number of executed operations (\#Op.), and energy consumption as our performance metrics.

To evaluate the performance, our hardware platform is the mobile Ampere GPU on Nvidia AGX Orin SoC with 5.33~TFLOPs, which is the flagship development board in AR/VR.
For GPU performance, we measure latency, including the execution time as well as the kernel launch on the mobile Ampere GPU. 
The GPU power is directly obtained using the built-in power sensing circuitry.

For comparison, we compare against three large-scale 3DGS algorithms: HierarchicalGS~\citep{kerbl2024hierarchical}, CityGaussian~\citep{liu2024citygaussian}, and Octree-GS~\citep{ren2024octree}.
We also compare against a dense 3DGS algorithm: 3DGS~\citep{kerbl20233d}.
In our evaluation, our algorithm \mode{\proj} applies optimizations in LOD search and splatting.

\subsection{Rendering Quality on Large-Scale Datasets}
\label{sec:eval:quality}

\begin{table*} 
\caption{Quantitative evaluation of \proj against the state-of-the-arts~\citep{kerbl20233d, liu2024citygaussian, ren2024octree, kerbl2024hierarchical}. Here, we show the results on three large-scale datasets: HierarchicalGS~\citep{kerbl2024hierarchical}, UrbanScene3D~\citep{lin2022capturing}, and MegaNeRF~\citep{turki2022mega}. We highlight the \goldtext{best} and \silvertext{second-best} results among all methods.}
\resizebox{\textwidth}{!}{
\renewcommand*{\arraystretch}{1}
\renewcommand*{\tabcolsep}{3pt}
\begin{tabular}{ c|cccccc|cccccc|cccccc } 
\toprule[0.15em]
Dataset & \multicolumn{6}{c|}{\textbf{HierarchicalGS}} & \multicolumn{6}{c|}{\textbf{UrbanScene3D}} & \multicolumn{6}{c}{\textbf{MegaNeRF}} \\ 
\midrule[0.05em]
\multirow{3}{*}{Metrics} & \multicolumn{3}{c|}{Quality} & \multicolumn{3}{c|}{Efficiency} & \multicolumn{3}{c|}{Quality} & \multicolumn{3}{c|}{Efficiency} & \multicolumn{3}{c|}{Quality} & \multicolumn{3}{c}{Efficiency} \\ 
& \specialcell{PSNR$\uparrow$\\(dB)} & SSIM$\uparrow$ &  \multicolumn{1}{r|}{LPIPS$\downarrow$}  & \specialcell{FPS$\uparrow$\\(Orin)}  & \specialcell{\#Op.$\downarrow$\\($10^6$)} & \specialcell{Energy$\downarrow$\\(mJ)} & \specialcell{PSNR$\uparrow$\\(dB)} & SSIM$\uparrow$ &  \multicolumn{1}{r|}{LPIPS$\downarrow$}  & \specialcell{FPS$\uparrow$\\(Orin)}  & \specialcell{\#Op.$\downarrow$\\($10^6$)} & \specialcell{Energy$\downarrow$\\(mJ)} & \specialcell{PSNR$\uparrow$\\(dB)} & SSIM$\uparrow$ &  \multicolumn{1}{r|}{LPIPS$\downarrow$}  & \specialcell{FPS$\uparrow$\\(Orin)}  & \specialcell{\#Op.$\downarrow$\\($10^6$)} & \specialcell{Energy$\downarrow$\\(mJ)} \\ 
\midrule[0.05em]
3DGS & 23.32 & 0.724 & 0.430 & 12.18 & 732.69 & 443.52 & 21.65 & \silvertext{0.818} & 0.269 & 11.60 & 2544.25 & 479.88 & 22.96 & 0.752 & 0.330 & 12.03 & 1630.92 & 415.46 \\
CityGaussian & 22.01 & 0.729 & 0.344 & 24.86 & 649.68 & 209.16 & 21.73 & \goldtext{0.825} & \goldtext{0.223} & 19.24 & 2522.30 & 276.16 & 24.27 & \goldtext{0.797} & \goldtext{0.238} & 20.40 & 2326.09 & 261.98 \\
OctreeGS & 21.33 & 0.646 & 0.415 & 13.55 & 2068.59 & 398.54 & 21.97 & 0.802 & 0.247 & 30.30 & 2121.52 & 186.43 & 24.56 & 0.765 & \silvertext{0.272} & 29.04 & 1198.70 & 176.87 \\
HierarchicalGS ($\tau$ = 3.0) & \goldtext{26.24} & \goldtext{0.812} & \goldtext{0.254} & 11.43 & 857.64 & 455.03 & \goldtext{23.79} & 0.755 & \silvertext{0.228} & 12.69 & 2502.38 & 409.96 & \goldtext{25.18} & \silvertext{0.783} & 0.280 & 15.39 & 1628.42 & 344.55 \\
HierarchicalGS ($\tau$ = 15.0) &  \silvertext{25.39} & \silvertext{0.771} & \silvertext{0.316} & 15.53 & 591.01 & 334.73 & 23.11 & 0.716 & 0.306 & 17.94 & 1547.26 & 290.29 & 23.31 & 0.686 & 0.367 & 19.63 & 1182.06 & 270.57 \\
\rowcolor{blue!6}
\proj ($\tau$ = 3.0) & \goldtext{26.24} & \goldtext{0.812} & \goldtext{0.254} & \silvertext{42.19} & \silvertext{584.15} & \silvertext{127.99} & \silvertext{23.78} & 0.755 & \silvertext{0.228} & \silvertext{39.01} & \silvertext{1118.84} & \silvertext{139.69} & \silvertext{25.17} & \silvertext{0.783} & 0.280 & \silvertext{56.42} &  \silvertext{528.81} & \silvertext{99.12} \\
\rowcolor{blue!6}
\proj ($\tau$ = 15.0) & \silvertext{25.39} & \silvertext{0.771} & \silvertext{0.316} & \goldtext{68.84} & \goldtext{363.68} & \goldtext{78.44} & 23.11 & 0.716 & 0.306 & \goldtext{58.67} & \goldtext{653.54} & \goldtext{92.05} & 23.31 & 0.686 & 0.367 & \goldtext{79.38} & \goldtext{362.02} & \goldtext{70.60} \\
\bottomrule[0.15em]
\end{tabular}
}
\label{tab:overall_tbl}
\end{table*}

\Tbl{tab:overall_tbl} shows the overall performance and quality comparison between \proj and other baselines.
Across three quality metrics, both \mode{HierarchicalGS} and \mode{\proj} with $\tau$ of 3 achieve the overall highest quality.
These two achieve similar scores on quality metrics, because the LUT-based exponential approximation in \Sect{sec:method:raster} introduces negligible accuracy loss.
Other optimizations introduced by \mode{\proj} all produce the bit-accurate results as \mode{HierarchicalGS}.
In comparison, dense models, e.g., 3DGS, perform poorly on large-scale scenes due to their lack of a mechanism to manage LoD.
More qualitative comparisons are shown in the appendix.

\begin{figure*}[t]
\centering
\begin{minipage}[t]{0.48\columnwidth}
  \centering
  \includegraphics[width=\columnwidth]{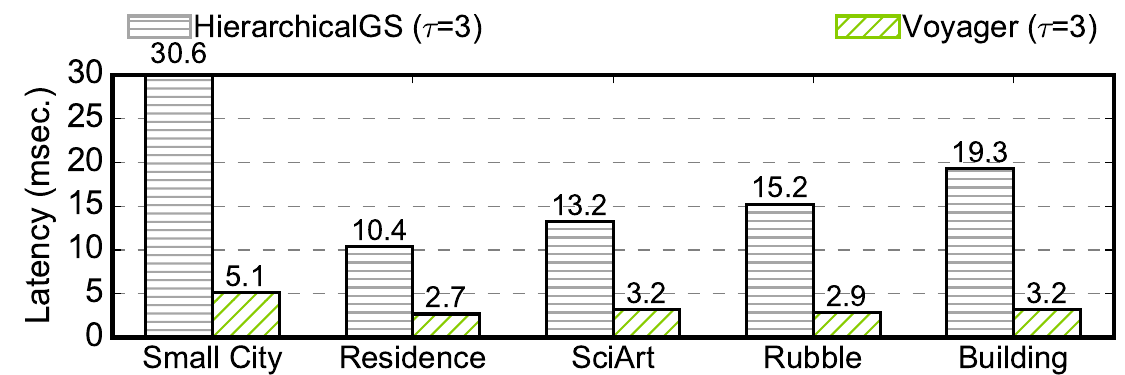}
  \caption{The standalone latency comparison of LoD search between \mode{HierarchicalGS} and \mode{\proj} with $\tau$ of 3. 
  We accelerate 5.1$\times$ speedup compared to \mode{HierarchicalGS}.
  }
  \label{fig:lod_speedup}
\end{minipage}
\hspace{2pt}
\begin{minipage}[t]{0.48\columnwidth}
  \centering
  \includegraphics[width=\columnwidth]{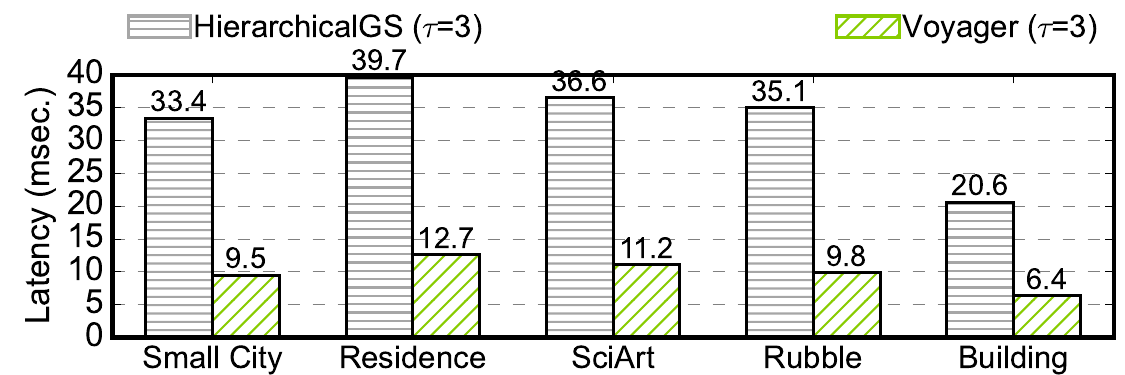}
  \caption{The standalone latency comparison of splatting between \mode{HierarchicalGS} and \mode{\proj} with $\tau$ of 3.
  We accelerate 3.3$\times$ speedup compared to \mode{HierarchicalGS}.
  }
  \label{fig:rasterization_speedup}
\end{minipage}
\end{figure*}

\subsection{Rendering Efficiency}
\label{sec:eval:perf}

\paragraph{Speedup.}
In terms of rendering speed, \mode{\proj} achieves the highest FPS among all evaluated methods in \Tbl{tab:overall_tbl}.
Specifically, \mode{\proj} can achieve up to 68.84, 58.67, and 79.38 FPS on three large-scale datasets, respectively.
It is also the only method capable of delivering real-time rendering ($\geq$60~FPS) consistently across all tested scenes. 
Compared to dense models such as 3DGS, \mode{\proj} delivers up to 6.6$\times$ speedup with better rendering quality. 
Compared to the corresponding large-scale 3DGS algorithm, \mode{HierarchicalGS}, \mode{\proj} still achieves 4.0$\times$ speedup.
\Fig{fig:lod_speedup} and \Fig{fig:rasterization_speedup} further dissect our speedup contribution and show the standalone speedup on LoD search and rasterization, respectively.
Compared to \mode{HierarchicalGS}, \mode{\proj} achieves 5.1$\times$ speedup on LoD search and 3.3$\times$ speedup on splatting (including projection, sorting, and rasterization).

In terms of total executed operations (\#Op.), \mode{\proj} achieves the lowest computational cost. 
This efficiency is primarily attributed to two key optimizations.
First, the temporal-aware LoD search reduces redundant tree node accesses.
Second, the preemptive $\alpha$-filtering shifts a transparency check to the early projection stage and naturally reduces the workload of the subsequent stages.
Lastly, the overall trend in energy savings follows the speedup results.
\mode{\proj} achieves the highest energy savings with 82\% and 80\%  of energy savings compared to \mode{3DGS} and \mode{HierarchicalGS}, repectively.

\subsection{Ablation Study}
\label{sec:eval:ablation}

\begin{table} 
\caption{Ablation Study of individual contributions in \proj on the HierarchicalGS dataset. 
FS: Fully-streaming LoD search. 
Temporal: Temporal-aware LoD search. FPS is measured on Orin.}
\centering
\resizebox{0.75\columnwidth}{!}{
\renewcommand*{\arraystretch}{1}
\renewcommand*{\tabcolsep}{3pt}
\begin{tabular}{ c|ccc|ccc } 
\toprule[0.15em]
\multirow{2}{*}{Metrics} & \multicolumn{3}{c|}{Quality} & \multicolumn{3}{c}{Efficiency}\\ 
& PSNR (dB)$\uparrow$ & SSIM$\uparrow$ &  \multicolumn{1}{r|}{LPIPS$\downarrow$}  & \multicolumn{1}{c}{FPS$\uparrow$} & \#Op. ($10^6$)$\downarrow$ & Energy (mJ)$\downarrow$ \\ 
\midrule[0.05em]
Base ($\tau$ = 3.0) & 26.24 & 0.812 & 0.254 & 11.43 & 857.64 & 455.03 \\
Base+FS ($\tau$ = 3.0) & 26.24 & 0.812 & 0.254 & 15.78 & 746.01 & 335.87 \\
Base+Temporal ($\tau$ = 3.0) & 26.24 & 0.812 & 0.254 & 22.15 & 681.63 & 239.12 \\
\proj ($\tau$ = 3.0) & 26.24 & 0.812 & 0.254 & 42.19 & 584.15 & 127.99 \\
\bottomrule[0.15em]
\end{tabular}
}
\label{tab:ablation_tbl}
\end{table}

\Tbl{tab:ablation_tbl} presents an ablation study that breaks down the contributions of individual components in \proj. 
Because the only approximation introduced by \proj is the LUT-based exponential approximation, it only introduces a negligible quality loss. 
All other acceleration techniques produce lossless and bit-accurate results.
From an efficiency perspective, each component contributes to the overall speedup.
Our fully streaming LoD search can achieve 1.4$\times$ speedup.
On top of that, our temporal-aware LoD search can further boost the overall performance to 1.9$\times$.
Due to Amdahl’s Law, only optimizing LoD search would not achieve a significant speedup.
By further accelerating the splatting stage via our preemptive $\alpha$-flitering, \proj eventually boosts the performance to 3.7$\times$ on the HierarchicalGS dataset.

\subsection{Evaluation on Small-Scale Datasets}
\label{sec:eval:small}

\begin{table} 
\caption{Evaluation of \proj against the state-of-the-arts on three small-scale datasets: Mip-NeRF360~\citep{barron2021mip}, Tank\&Temple~\citep{knapitsch2017}, DeepBlending~\citep{hedman2018deep}. We highlight the \goldtext{best} and \silvertext{second-best} results among all methods. \textit{Note that, rendering small-scale scenes is \textbf{not} the primary focus of this work; we include the results for completeness.}}
\resizebox{\columnwidth}{!}{
\renewcommand*{\arraystretch}{1}
\renewcommand*{\tabcolsep}{3pt}
\begin{tabular}{ c|ccccc|ccccc|ccccc } 
\toprule[0.15em]
Dataset & \multicolumn{5}{c|}{\textbf{MipNeRF360}} & \multicolumn{5}{c|}{\textbf{Tank\&Temple}} & \multicolumn{5}{c}{\textbf{DeepBlending}} \\ 
\midrule[0.05em]
\multirow{3}{*}{Metrics} & \multicolumn{3}{c|}{Quality} & \multicolumn{2}{c|}{Efficiency} & \multicolumn{3}{c|}{Quality} & \multicolumn{2}{c|}{Efficiency} & \multicolumn{3}{c|}{Quality} & \multicolumn{2}{c}{Efficiency} \\ 
& \specialcell{PSNR$\uparrow$\\(dB)} & SSIM$\uparrow$ &  \multicolumn{1}{r|}{LPIPS$\downarrow$}  & \specialcell{FPS$\uparrow$\\(Orin)}  &  \specialcell{Energy$\downarrow$\\(mJ)} & \specialcell{PSNR$\uparrow$\\(dB)} & SSIM$\uparrow$ &  \multicolumn{1}{r|}{LPIPS$\downarrow$}  & \specialcell{FPS$\uparrow$\\(Orin)}  & \specialcell{Energy$\downarrow$\\(mJ)} & \specialcell{PSNR$\uparrow$\\(dB)} & SSIM$\uparrow$ &  \multicolumn{1}{r|}{LPIPS$\downarrow$}  & \specialcell{FPS$\uparrow$\\(Orin)}  &  \specialcell{Energy$\downarrow$\\(mJ)} \\ 
\midrule[0.05em]
3DGS & \silvertext{27.53} & 0.815 & 0.220 & 13.07 & 428.79 & 23.76 & \silvertext{0.852} & \goldtext{0.169} & 23.95 & 215.21 & 29.80 & \goldtext{0.907} & \goldtext{0.238} & 14.73  & 368.21 \\
CityGaussian & 27.50 & 0.813 & 0.221 & 14.42 & 380.82 & 23.71 & 0.848 & 0.177 & 27.20 & 193.11  &  29.54 & \silvertext{0.904} & \silvertext{0.244} & 14.63 & 382.83 \\
OctreeGS & \goldtext{27.80} & \goldtext{0.846} & \goldtext{0.198} & 20.16 & 362.75 & 24.39 & \goldtext{0.856} & \silvertext{0.170} & 36.31 & 135.10 & \silvertext{29.92} & 0.903 & 0.262 & 26.67  & 141.18 \\
HierarchicalGS ($\tau$ = 3.0) & 27.24 & \silvertext{0.818} & \silvertext{0.213} & 16.44 & 320.30 & \goldtext{25.65} & 0.850 & 0.170 & 24.93 & 194.06 & \goldtext{30.29} & 0.904 & 0.260 & 26.95  & 203.76 \\
HierarchicalGS ($\tau$ = 15.0) & 24.71 & 0.718 & 0.294 & 16.89 & 308.00 & \silvertext{24.72} & 0.819 & 0.233 & 31.95 & 149.19 & 29.02 & 0.892 & 0.277 & 28.71 & 191.97 \\
\rowcolor{blue!6}
\proj ($\tau$ = 3.0) & 27.24 & \silvertext{0.818} & \silvertext{0.213} & \silvertext{58.10} & \silvertext{110.38} & \goldtext{25.65} & 0.850 & 0.170 & \silvertext{46.52} & \silvertext{134.30} & \goldtext{30.29} & 0.904 & 0.260 & \silvertext{53.27} & \silvertext{131.44} \\
\rowcolor{blue!6}
\proj ($\tau$ = 15.0) & 24.71 & 0.718 & 0.294 & \goldtext{70.28} & \goldtext{90.04} & \silvertext{24.72} & 0.819 & 0.233 & \goldtext{93.45} & \goldtext{66.97} & 29.02 & 0.892 & 0.277 & \goldtext{67.91} & \goldtext{103.13} \\
\bottomrule[0.15em]
\end{tabular}
    }
\label{tab:small_scene_tbl}
\end{table}

We also evaluate \proj on small-scale datasets in \Tbl{tab:small_scene_tbl}.
While rendering small-scale scenes is not the primary focus of this work, we include the results for completeness.
We show that \mode{\proj} achieves relatively low rendering quality compared to the corresponding \mode{HierarchicalGS}.
The primary reason is that the LUT approximation introduces some artifacts compared to the exact $\alpha$-computation.
Nevertheless, \mode{\proj} can achieve 2.5$\times$ and 3.4$\times$ speedup against \mode{HierarchicalGS} and \mode{3DGS}, respectively.
With some degree of quality sacrifices, \mode{\proj} at $\tau$ of 15 can further boost the performance to 3.6$\times$ and 4.8$\times$ speedup, respectively.
This could be a potential limitation of \proj on small-scale datasets, which we leave for future work.

\subsection{Senstivity Study}

\begin{figure*}[t]
\centering
\subfloat[Small City.]{
	\label{fig:lut_smallcity}	
        \includegraphics[width=0.195\columnwidth]{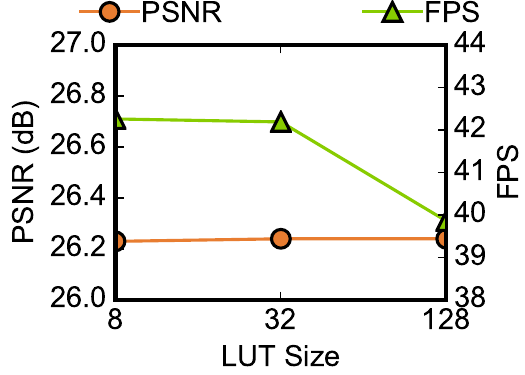}}
\subfloat[Residence.]{
	\label{fig:lut_residence}
	\includegraphics[width=0.195\columnwidth]{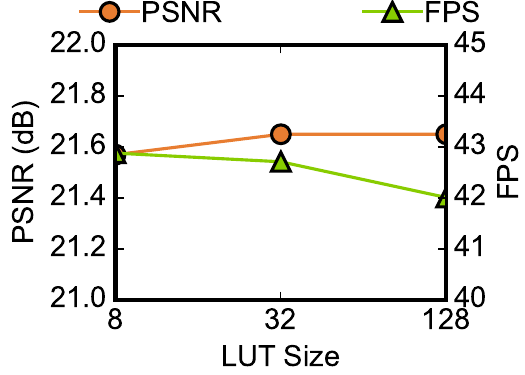}} 
\subfloat[SciArt.]{
	\label{fig:lut_sciart}	
        \includegraphics[width=0.195\columnwidth]{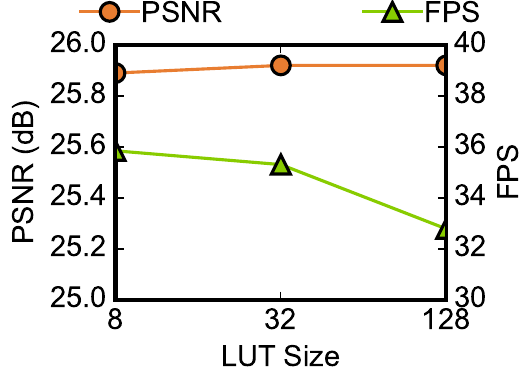}}
\subfloat[Rubble.]{
	\label{fig:lut_rubble}
	\includegraphics[width=0.195\columnwidth]{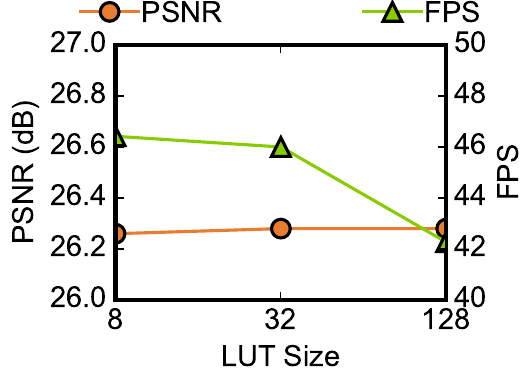}} 
\subfloat[Building.]{
	\label{fig:lut_building}	
        \includegraphics[width=0.195\columnwidth]{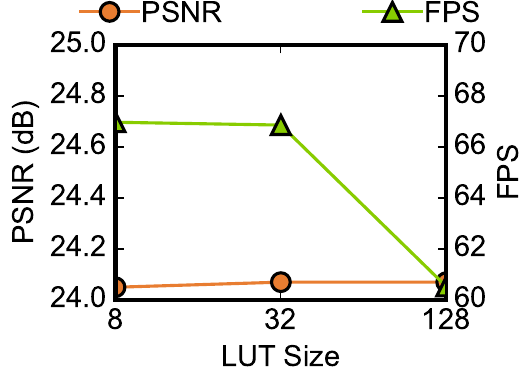}}
\caption{The sensitivity of rendering quality and performance to the lookup table size ($m$ intervals) in the rasterization stage. As $m$ increases, the performance decreases while the accuracy increases a little. We found that the size of $32$ is a good trade-off across different datasets.}
\label{fig:lut_size}
\end{figure*}

\Fig{fig:lut_size} shows the sensitivity study of rendering quality and the performance to the LUT size ($m$ intervals in \Sect{sec:method:raster}), across five different scenes.
We observe that increasing the LUT size from 8 to 32 results in notable improvements in rendering quality, while incurring only minimal performance overhead.
However, further increasing the LUT size from 32 to 128 leads to a significant drop in performance, with negligible gains in visual quality.
This trade-off suggests that a LUT size of 32 strikes a good balance between performance and rendering fidelity.
\section{Conclusion}
\label{sec:conc}

Human imagination is boundless.
When harnessing Gaussians to construct virtual worlds, our rendering systems should be just as limitless.
\proj marks the first attempt toward enabling real-time city-scale Gaussian splatting on resource-constrained devices.
By identifying and addressing the two primary bottlenecks, LoD search and splatting, we introduce two key innovations:
temporal-aware LoD search and preemptive $\alpha$-filtering, both of which significantly enhance rendering efficiency.
Overall, \proj achieves up to 6.6$\times$ speedup in large-scale scenes and 4.8$\times$ speedup in small-scale scenes, all with comparable rendering quality.


\bibliographystyle{iclr2026_conference}
\bibliography{references}

\begin{thebibliography}{41}
\providecommand{\natexlab}[1]{#1}
\providecommand{\url}[1]{\texttt{#1}}
\expandafter\ifx\csname urlstyle\endcsname\relax
  \providecommand{\doi}[1]{doi: #1}\else
  \providecommand{\doi}{doi: \begingroup \urlstyle{rm}\Url}\fi

\bibitem[Amdahl(1967)]{amdahl1967validity}
Gene~M Amdahl.
\newblock Validity of the single processor approach to achieving large scale computing capabilities.
\newblock In \emph{Proceedings of the April 18-20, 1967, spring joint computer conference}, pp.\  483--485, 1967.

\bibitem[Barron et~al.(2021)Barron, Mildenhall, Tancik, Hedman, Martin-Brualla, and Srinivasan]{barron2021mip}
Jonathan~T Barron, Ben Mildenhall, Matthew Tancik, Peter Hedman, Ricardo Martin-Brualla, and Pratul~P Srinivasan.
\newblock Mip-nerf: A multiscale representation for anti-aliasing neural radiance fields.
\newblock In \emph{Proceedings of the IEEE/CVF International Conference on Computer Vision}, pp.\  5855--5864, 2021.

\bibitem[Barron et~al.(2023)Barron, Mildenhall, Verbin, Srinivasan, and Hedman]{barron2023zip}
Jonathan~T Barron, Ben Mildenhall, Dor Verbin, Pratul~P Srinivasan, and Peter Hedman.
\newblock Zip-nerf: Anti-aliased grid-based neural radiance fields.
\newblock In \emph{Proceedings of the IEEE/CVF International Conference on Computer Vision}, pp.\  19697--19705, 2023.

\bibitem[Bauer et~al.(2014)Bauer, Treichler, and Aiken]{bauer2014singe}
Michael Bauer, Sean Treichler, and Alex Aiken.
\newblock Singe: Leveraging warp specialization for high performance on gpus.
\newblock In \emph{Proceedings of the 19th ACM SIGPLAN symposium on Principles and practice of parallel programming}, pp.\  119--130, 2014.

\bibitem[Dai et~al.(2024)Dai, Xu, Xie, Liu, Wang, and Xu]{dai2024high}
Pinxuan Dai, Jiamin Xu, Wenxiang Xie, Xinguo Liu, Huamin Wang, and Weiwei Xu.
\newblock High-quality surface reconstruction using gaussian surfels.
\newblock In \emph{ACM SIGGRAPH 2024 Conference Papers}, pp.\  1--11, 2024.

\bibitem[Fan et~al.(2024)Fan, Wang, Wen, Zhu, Xu, Wang, et~al.]{fan2023lightgaussian}
Zhiwen Fan, Kevin Wang, Kairun Wen, Zehao Zhu, Dejia Xu, Zhangyang Wang, et~al.
\newblock Lightgaussian: Unbounded 3d gaussian compression with 15x reduction and 200+ fps.
\newblock \emph{Advances in neural information processing systems}, 37:\penalty0 140138--140158, 2024.

\bibitem[Fang \& Wang(2024)Fang and Wang]{fang2024mini}
Guangchi Fang and Bing Wang.
\newblock Mini-splatting: Representing scenes with a constrained number of gaussians.
\newblock In \emph{European Conference on Computer Vision}, pp.\  165--181. Springer, 2024.

\bibitem[Feng et~al.(2024{\natexlab{a}})Feng, Chen, Fu, Liao, Wang, Liu, Pei, Li, Zhang, and Dai]{feng2024flashgs}
Guofeng Feng, Siyan Chen, Rong Fu, Zimu Liao, Yi~Wang, Tao Liu, Zhilin Pei, Hengjie Li, Xingcheng Zhang, and Bo~Dai.
\newblock Flashgs: Efficient 3d gaussian splatting for large-scale and high-resolution rendering.
\newblock \emph{arXiv preprint arXiv:2408.07967}, 2024{\natexlab{a}}.

\bibitem[Feng et~al.(2024{\natexlab{b}})Feng, Cao, Chen, Xu, Mu, Martin, and Hu]{feng2024evsplitting}
Qi-Yuan Feng, Geng-Chen Cao, Hao-Xiang Chen, Qun-Ce Xu, Tai-Jiang Mu, Ralph Martin, and Shi-Min Hu.
\newblock Evsplitting: an efficient and visually consistent splitting algorithm for 3d gaussian splatting.
\newblock In \emph{SIGGRAPH Asia 2024 Conference Papers}, pp.\  1--11, 2024{\natexlab{b}}.

\bibitem[Franke et~al.(2024)Franke, Fink, and Stamminger]{franke2024vr}
Linus Franke, Laura Fink, and Marc Stamminger.
\newblock Vr-splatting: Foveated radiance field rendering via 3d gaussian splatting and neural points.
\newblock \emph{arXiv preprint arXiv:2410.17932}, 2024.

\bibitem[Girish et~al.(2024)Girish, Gupta, and Shrivastava]{girish2024eagles}
Sharath Girish, Kamal Gupta, and Abhinav Shrivastava.
\newblock Eagles: Efficient accelerated 3d gaussians with lightweight encodings.
\newblock In \emph{European Conference on Computer Vision}, pp.\  54--71. Springer, 2024.

\bibitem[Google(2025)]{ortool}
Google.
\newblock Or-tools - google optimization tools.
\newblock \url{https://github.com/google/or-tools}, 2025.

\bibitem[Gui et~al.(2024)Gui, Hu, Chen, Huang, Yin, Yang, and Wu]{gui2024balanced}
Hao Gui, Lin Hu, Rui Chen, Mingxiao Huang, Yuxin Yin, Jin Yang, and Yong Wu.
\newblock Balanced 3dgs: Gaussian-wise parallelism rendering with fine-grained tiling.
\newblock \emph{arXiv preprint arXiv:2412.17378}, 2024.

\bibitem[Hahlbohm et~al.(2025)Hahlbohm, Friederichs, Weyrich, Franke, Kappel, Castillo, Stamminger, Eisemann, and Magnor]{hahlbohm2025efficient}
Florian Hahlbohm, Fabian Friederichs, Tim Weyrich, Linus Franke, Moritz Kappel, Susana Castillo, Marc Stamminger, Martin Eisemann, and Marcus Magnor.
\newblock Efficient perspective-correct 3d gaussian splatting using hybrid transparency.
\newblock In \emph{Computer Graphics Forum}, pp.\  e70014. Wiley Online Library, 2025.

\bibitem[Hedman et~al.(2018)Hedman, Philip, Price, Frahm, Drettakis, and Brostow]{hedman2018deep}
Peter Hedman, Julien Philip, True Price, Jan-Michael Frahm, George Drettakis, and Gabriel Brostow.
\newblock Deep blending for free-viewpoint image-based rendering.
\newblock \emph{ACM Transactions on Graphics (ToG)}, 37\penalty0 (6):\penalty0 1--15, 2018.

\bibitem[Huang et~al.(2024)Huang, Yu, Chen, Geiger, and Gao]{huang20242d}
Binbin Huang, Zehao Yu, Anpei Chen, Andreas Geiger, and Shenghua Gao.
\newblock 2d gaussian splatting for geometrically accurate radiance fields.
\newblock In \emph{ACM SIGGRAPH 2024 conference papers}, pp.\  1--11, 2024.

\bibitem[Huang et~al.(2025)Huang, Zhu, Liu, Lin, Liu, He, Leng, Guo, and Feng]{huang2025seele}
Xiaotong Huang, He~Zhu, Zihan Liu, Weikai Lin, Xiaohong Liu, Zhezhi He, Jingwen Leng, Minyi Guo, and Yu~Feng.
\newblock Seele: A unified acceleration framework for real-time gaussian splatting.
\newblock \emph{arXiv preprint arXiv:2503.05168}, 2025.

\bibitem[Kerbl et~al.(2023)Kerbl, Kopanas, Leimk{\"u}hler, and Drettakis]{kerbl20233d}
Bernhard Kerbl, Georgios Kopanas, Thomas Leimk{\"u}hler, and George Drettakis.
\newblock 3d gaussian splatting for real-time radiance field rendering.
\newblock \emph{ACM Transactions on Graphics}, 42\penalty0 (4):\penalty0 1--14, 2023.

\bibitem[Kerbl et~al.(2024)Kerbl, Meuleman, Kopanas, Wimmer, Lanvin, and Drettakis]{kerbl2024hierarchical}
Bernhard Kerbl, Andreas Meuleman, Georgios Kopanas, Michael Wimmer, Alexandre Lanvin, and George Drettakis.
\newblock A hierarchical 3d gaussian representation for real-time rendering of very large datasets.
\newblock \emph{ACM Transactions on Graphics (TOG)}, 43\penalty0 (4):\penalty0 1--15, 2024.

\bibitem[Knapitsch et~al.(2017)Knapitsch, Park, Zhou, and Koltun]{knapitsch2017}
Arno Knapitsch, Jaesik Park, Qian-Yi Zhou, and Vladlen Koltun.
\newblock Tanks and temples: Benchmarking large-scale scene reconstruction.
\newblock \emph{ACM Transactions on Graphics}, 36\penalty0 (4), 2017.

\bibitem[Krashinsky et~al.(2020)Krashinsky, Giroux, Jones, Stam, and Ramaswamy]{ampere_arch}
Ronny Krashinsky, Olivier Giroux, Stephen Jones, Nick Stam, and Sridhar Ramaswamy.
\newblock Nvidia ampere architecture in-depth, 2020.
\newblock URL \url{https://developer.nvidia.com/blog/nvidia-ampere-architecture-in-depth/}.

\bibitem[Lee et~al.(2024{\natexlab{a}})Lee, Rho, Sun, Ko, and Park]{lee2024compact}
Joo~Chan Lee, Daniel Rho, Xiangyu Sun, Jong~Hwan Ko, and Eunbyung Park.
\newblock Compact 3d gaussian representation for radiance field.
\newblock In \emph{Proceedings of the IEEE/CVF Conference on Computer Vision and Pattern Recognition}, pp.\  21719--21728, 2024{\natexlab{a}}.

\bibitem[Lee et~al.(2024{\natexlab{b}})Lee, Lee, Lee, Park, and Sim]{lee2024gscore}
Junseo Lee, Seokwon Lee, Jungi Lee, Junyong Park, and Jaewoong Sim.
\newblock Gscore: Efficient radiance field rendering via architectural support for 3d gaussian splatting.
\newblock In \emph{Proceedings of the 29th ACM International Conference on Architectural Support for Programming Languages and Operating Systems, Volume 3}, pp.\  497--511, 2024{\natexlab{b}}.

\bibitem[Lin et~al.(2022)Lin, Liu, Hu, Yan, Xie, and Huang]{lin2022capturing}
Liqiang Lin, Yilin Liu, Yue Hu, Xingguang Yan, Ke~Xie, and Hui Huang.
\newblock Capturing, reconstructing, and simulating: the urbanscene3d dataset.
\newblock In \emph{European Conference on Computer Vision}, pp.\  93--109. Springer, 2022.

\bibitem[Lin et~al.(2025)Lin, Feng, and Zhu]{lin2025metasapiens}
Weikai Lin, Yu~Feng, and Yuhao Zhu.
\newblock Metasapiens: Real-time neural rendering with efficiency-aware pruning and accelerated foveated rendering.
\newblock In \emph{Proceedings of the 30th ACM International Conference on Architectural Support for Programming Languages and Operating Systems, Volume 1}, pp.\  669--682, 2025.

\bibitem[Liu et~al.(2024)Liu, Luo, Fan, Wang, Peng, and Zhang]{liu2024citygaussian}
Yang Liu, Chuanchen Luo, Lue Fan, Naiyan Wang, Junran Peng, and Zhaoxiang Zhang.
\newblock Citygaussian: Real-time high-quality large-scale scene rendering with gaussians.
\newblock In \emph{European Conference on Computer Vision}, pp.\  265--282. Springer, 2024.

\bibitem[Mallick et~al.(2024)Mallick, Goel, Kerbl, Steinberger, Carrasco, and De~La~Torre]{mallick2024taming}
Saswat~Subhajyoti Mallick, Rahul Goel, Bernhard Kerbl, Markus Steinberger, Francisco~Vicente Carrasco, and Fernando De~La~Torre.
\newblock Taming 3dgs: High-quality radiance fields with limited resources.
\newblock In \emph{SIGGRAPH Asia 2024 Conference Papers}, pp.\  1--11, 2024.

\bibitem[Martello \& Toth(1990)Martello and Toth]{martello1990knapsack}
Silvano Martello and Paolo Toth.
\newblock \emph{Knapsack problems: algorithms and computer implementations}.
\newblock John Wiley \& Sons, Inc., 1990.

\bibitem[Niemeyer et~al.(2024)Niemeyer, Manhardt, Rakotosaona, Oechsle, Duckworth, Gosula, Tateno, Bates, Kaeser, and Tombari]{niemeyer2024radsplat}
Michael Niemeyer, Fabian Manhardt, Marie-Julie Rakotosaona, Michael Oechsle, Daniel Duckworth, Rama Gosula, Keisuke Tateno, John Bates, Dominik Kaeser, and Federico Tombari.
\newblock Radsplat: Radiance field-informed gaussian splatting for robust real-time rendering with 900+ fps.
\newblock \emph{arXiv preprint arXiv:2403.13806}, 2024.

\bibitem[Nvidia(2023)]{orinsoc}
Nvidia.
\newblock Jetson orin for next-gen robotics, 2023.
\newblock URL \url{https://www.nvidia.com/en-us/autonomous-machines/embedded-systems/jetson-orin/}.

\bibitem[Nvidia(2025)]{nsight}
Nvidia.
\newblock Nvidia nsight compute, 2025.
\newblock URL \url{https://docs.nvidia.com/nsight-compute/ProfilingGuide/index.html#metrics-reference}.

\bibitem[Peng et~al.(2024)Peng, Shao, Liu, Zhou, Yang, Wang, and Zhou]{peng2024rtg}
Zhexi Peng, Tianjia Shao, Yong Liu, Jingke Zhou, Yin Yang, Jingdong Wang, and Kun Zhou.
\newblock Rtg-slam: Real-time 3d reconstruction at scale using gaussian splatting.
\newblock In \emph{ACM SIGGRAPH 2024 Conference Papers}, pp.\  1--11, 2024.

\bibitem[Radl et~al.(2024)Radl, Steiner, Parger, Weinrauch, Kerbl, and Steinberger]{radl2024stopthepop}
Lukas Radl, Michael Steiner, Mathias Parger, Alexander Weinrauch, Bernhard Kerbl, and Markus Steinberger.
\newblock Stopthepop: Sorted gaussian splatting for view-consistent real-time rendering.
\newblock \emph{ACM Transactions on Graphics (TOG)}, 43\penalty0 (4):\penalty0 1--17, 2024.

\bibitem[Ren et~al.(2024)Ren, Jiang, Lu, Yu, Xu, Ni, and Dai]{ren2024octree}
Kerui Ren, Lihan Jiang, Tao Lu, Mulin Yu, Linning Xu, Zhangkai Ni, and Bo~Dai.
\newblock Octree-gs: Towards consistent real-time rendering with lod-structured 3d gaussians.
\newblock \emph{arXiv preprint arXiv:2403.17898}, 2024.

\bibitem[Roessle et~al.(2024)Roessle, M{\"u}ller, Porzi, Rota~Bul{\`o}, Kontschieder, Dai, and Nie{\ss}ner]{roessle2024l3dg}
Barbara Roessle, Norman M{\"u}ller, Lorenzo Porzi, Samuel Rota~Bul{\`o}, Peter Kontschieder, Angela Dai, and Matthias Nie{\ss}ner.
\newblock L3dg: Latent 3d gaussian diffusion.
\newblock In \emph{SIGGRAPH Asia 2024 Conference Papers}, pp.\  1--11, 2024.

\bibitem[Tu et~al.(2024)Tu, Kerbl, and de~la Torre]{tu2024fast}
Xuechang Tu, Bernhard Kerbl, and Fernando de~la Torre.
\newblock Fast and robust 3d gaussian splatting for virtual reality.
\newblock In \emph{SIGGRAPH Asia 2024 Posters}, pp.\  1--3. 2024.

\bibitem[Turki et~al.(2022)Turki, Ramanan, and Satyanarayanan]{turki2022mega}
Haithem Turki, Deva Ramanan, and Mahadev Satyanarayanan.
\newblock Mega-nerf: Scalable construction of large-scale nerfs for virtual fly-throughs.
\newblock In \emph{Proceedings of the IEEE/CVF conference on computer vision and pattern recognition}, pp.\  12922--12931, 2022.

\bibitem[Turki et~al.(2023)Turki, Zollh{\"o}fer, Richardt, and Ramanan]{turki2023pynerf}
Haithem Turki, Michael Zollh{\"o}fer, Christian Richardt, and Deva Ramanan.
\newblock Pynerf: Pyramidal neural radiance fields.
\newblock \emph{Advances in neural information processing systems}, 36:\penalty0 37670--37681, 2023.

\bibitem[Wang et~al.(2024{\natexlab{a}})Wang, Yi, and Ma]{wang2024adr}
Xinzhe Wang, Ran Yi, and Lizhuang Ma.
\newblock Adr-gaussian: Accelerating gaussian splatting with adaptive radius.
\newblock In \emph{SIGGRAPH Asia 2024 Conference Papers}, pp.\  1--10, 2024{\natexlab{a}}.

\bibitem[Wang et~al.(2024{\natexlab{b}})Wang, Deng, Zhang, Jakob, and Marschner]{wang2024simple}
Zichen Wang, Xi~Deng, Ziyi Zhang, Wenzel Jakob, and Steve Marschner.
\newblock A simple approach to differentiable rendering of sdfs.
\newblock In \emph{SIGGRAPH Asia 2024 Conference Papers}, pp.\  1--11, 2024{\natexlab{b}}.

\bibitem[Xu et~al.(2023)Xu, Agrawal, Laney, Garcia, Bansal, Kim, Rota~Bul{\`o}, Porzi, Kontschieder, Bo{\v{z}}i{\v{c}}, et~al.]{xu2023vr}
Linning Xu, Vasu Agrawal, William Laney, Tony Garcia, Aayush Bansal, Changil Kim, Samuel Rota~Bul{\`o}, Lorenzo Porzi, Peter Kontschieder, Alja{\v{z}} Bo{\v{z}}i{\v{c}}, et~al.
\newblock Vr-nerf: High-fidelity virtualized walkable spaces.
\newblock In \emph{SIGGRAPH Asia 2023 Conference Papers}, pp.\  1--12, 2023.

\end{thebibliography}

\clearpage

\begin{alphasection}

\section{Supplementary}
\label{sec:supple}

\subsection{Experimental Setup}

\paragraph{Hardware Platforms.}
We conduct both the performance and accuracy measurements on Nvidia Jetson Orin SoC with 5.325 TFLOPS (FP32) and 64 GB memory.

\paragraph{Evaluation Metrics.}
Next, we describe how we obtain various performance and quality metrics.
\begin{itemize}
    \item \textbf{Image Consistency (PSNR, SSIM, LPIPS).} To assess frame-to-frame image consistency, we compared generated videos by different acceleration methods against the original video results on image quality, using PSNR, SSIM, and LPIPS.
    PSNR is calculated using standard image processing libraries, e.g.,\texttt{skimage.metrics.peak\_signal\_noise\_ratio}.
    Similarly, SSIM is calculated using  \texttt{skimage.metrics.structural\_similarity}.
    LPIPS is measured using the \texttt{lpips} library in Python.
    \item \textbf{Performance.} 
    For performance, we report end-to-end latency, GPU memory consumption, and GPU energy consumption.
    For end-to-end latency, we measure the total time elapsed during one single frame rendering.
    Here, we use Python's \texttt{torch.cuda} module. 
    We capture the start and end timestamps using \texttt{torch.cuda.event()} and use the difference between these two as the end-to-end latency. 
    For GPU memory consumption, we use the built-in measurement to monitor the peak GPU memory usage during inference.
    The GPU power is directly obtained using the built-in power sensing circuitry on Orin.
\end{itemize}

\begin{table*} 
\caption{Efficiency evaluation of \proj against the state-of-the-arts~\citep{kerbl20233d, lee2024compact, liu2024citygaussian, ren2024octree, kerbl2024hierarchical}. Here, we show the results on three large-scale datasets: HierarchicalGS~\citep{kerbl2024hierarchical}, UrbanScene3D~\citep{lin2022capturing}, and MegaNeRF~\citep{turki2022mega}.}
\resizebox{\textwidth}{!}{
\renewcommand*{\arraystretch}{1}
\renewcommand*{\tabcolsep}{5pt}
\begin{tabular}{ c|c|c|ccc|ccc } 
\toprule[0.15em]
\multirow{2}{*}{\textbf{Quality Metrics}} & \multirow{2}{*}{\textbf{Methods}} & \multicolumn{1}{c|}{\textbf{HierarchicalGS}} & \multicolumn{3}{c|}{\textbf{UrbanScene3D}} & \multicolumn{3}{c}{\textbf{MegaNeRF}} \\ 
 & & \multicolumn{1}{c|}{Small City} & Residence & SciArt & \multicolumn{1}{c|}{Average} & Rubble & Building & Average \\ 
\midrule[0.05em]
\multirow{9}{*}{FPS (Orin) $\uparrow$} & 3DGS &12.18 &13.61 &9.59 &11.60 &12.28 &11.78 &12.03 \\
& CityGaussian &24.86 &18.47 &20.00 &19.24 &25.20 &15.59 &20.40 \\
& OctreeGS &13.55 &32.63 & 27.97 & 30.30 &31.24 &26.85 &29.04 \\
& HierarchicalGS ($\tau$ = 0.0) &8.95 &13.02 &11.70 &12.36 &12.70 &16.97 &14.83 \\
& HierarchicalGS ($\tau$ = 3.0) &11.43 &12.93 &12.44 &12.69 &13.25 &17.52 &15.39 \\
& HierarchicalGS ($\tau$ = 15.0) &15.53 &17.25 &18.63 &17.94 &16.79 &22.46 &19.63 \\
& \proj ($\tau$ = 0.0) &28.70 &42.24 &32.63 &37.43 &43.55 &63.65 &53.60 \\
& \proj ($\tau$ = 3.0) &42.19 &42.71 &35.31 &39.01 &46.0 &66.86 &56.42 \\
& \proj ($\tau$ = 15.0) &68.84 &58.19 &59.15 &58.67 &64.24 &94.52 &79.38 \\
\midrule[0.05em]
\multirow{9}{*}{\#Op. ($10^6$)$\downarrow$} & 3DGS &732.69 &2225.49 &2863.02 &2544.25 &1407.22 &1854.63 &1630.92 \\
& CityGaussian &649.68 &2473.56 &2571.05 &2522.30 &1177.23 &3474.94 &2326.09 \\
& OctreeGS &2068.59 &2088.34 &2154.70 &2121.52 &1389.11 &1008.29 &1198.70 \\
& HierarchicalGS ($\tau$ = 0.0) &1086.99 &2338.85 &2946.90 &2642.88 &1993.58 &1359.21 &1676.39 \\
& HierarchicalGS ($\tau$ = 3.0) &857.64 &2323.27 &2681.50 &2502.38 &1948.93 &1307.92 &1628.42 \\
& HierarchicalGS ($\tau$ = 15.0) &591.01 &1587.84 &1506.68 &1547.26 &1368.09 &996.03 &1182.06 \\
& \proj ($\tau$ = 0.0) &682.10 &1205.48 &1180.87 &1193.17 &580.58 &522.95 &551.77 \\
& \proj ($\tau$ = 3.0) &584.15 &1166.39 &1071.29 &1118.84 &556.27 &501.35 &528.81 \\
& \proj ($\tau$ = 15.0) &363.67 &712.98 &594.10 &653.54 &391.20 &332.85 &362.02 \\
\midrule[0.05em]
\multirow{9}{*}{Energy (mJ)$\downarrow$} & 3DGS &443.52 &396.78 &562.99 &479.88 &423.37 &407.54 &415.46 \\
& CityGaussian &209.16 & 292.30 & 260.01 & 276.16 & 190.47 & 333.49 & 261.98 \\
& OctreeGS &398.54 & 207.37 & 165.49 & 186.43 & 160.04 & 193.69 & 176.87 \\
& HierarchicalGS ($\tau$ = 0.0) &581.16 & 399.37 & 444.44 & 421.90 & 409.61 & 306.37 & 357.99 \\
& HierarchicalGS ($\tau$ = 3.0) &455.03 & 402.08 & 417.85 & 409.96 & 392.37 & 296.73 & 344.55 \\
& HierarchicalGS ($\tau$ = 15.0) &405.95 & 378.15 & 371.99 & 375.07 & 372.24 & 273.68 & 322.96 \\
& \proj ($\tau$ = 0.0) &188.13 & 127.85 & 165.50 & 146.67 & 123.99 & 84.83 & 104.41 \\
& \proj ($\tau$ = 3.0) &127.99 & 126.44 & 152.94 & 139.69 & 117.47 & 80.76 & 99.12 \\
& \proj ($\tau$ = 15.0) &78.44 & 92.80 & 91.29 & 92.05 & 84.06 & 57.13 & 70.60 \\
\bottomrule[0.15em]
\end{tabular}
}
\label{tab:overall_tbl2}
\end{table*}

\begin{table*} 
\caption{Quality evaluation of \proj against the state-of-the-arts~\citep{kerbl20233d, lee2024compact, liu2024citygaussian, ren2024octree, kerbl2024hierarchical}. Here, we show the results on three large-scale datasets: HierarchicalGS~\citep{kerbl2024hierarchical}, UrbanScene3D~\citep{lin2022capturing}, and MegaNeRF~\citep{turki2022mega}. We do not report the result of \textit{Campus} due to OOM on all methods.}
\resizebox{\textwidth}{!}{
\renewcommand*{\arraystretch}{1}
\renewcommand*{\tabcolsep}{5pt}
\begin{tabular}{ c|c|c|ccc|ccc } 
\toprule[0.15em]
\multirow{2}{*}{\textbf{Quality Metrics}} & \multirow{2}{*}{\textbf{Methods}} & \multicolumn{1}{c|}{\textbf{HierarchicalGS}} & \multicolumn{3}{c|}{\textbf{UrbanScene3D}} & \multicolumn{3}{c}{\textbf{MegaNeRF}} \\ 
 & & \multicolumn{1}{c|}{Small City} & Residence & SciArt & \multicolumn{1}{c|}{Average} & Rubble & Building & Average \\ 
\midrule[0.05em]
\multirow{9}{*}{PSNR (dB) $\uparrow$} & 3DGS &23.32 &21.86 &21.43 &21.65 &25.68 &20.25 &22.96 \\
& CityGaussian &22.01 &22.03 &21.43 &21.73 &25.59 &22.96 &24.27 \\
& OctreeGS &21.33 &21.77 &22.18 &21.97 &25.56 &23.55 &24.56 \\
& HierarchicalGS ($\tau$ = 0.0) &26.34 &21.69 &25.95 &23.82 &26.47 &24.15 &25.31 \\
& HierarchicalGS ($\tau$ = 3.0) &26.24 &21.65 &25.92 &23.79 &26.29 &24.07 &25.18 \\
& HierarchicalGS ($\tau$ = 15.0) &25.39 &21.11 &25.11 &23.11 &23.88 &22.75 &23.31 \\
& \proj ($\tau$ = 0.0) &26.34 &21.69 &25.95 &23.82 &26.47 &24.15 &25.31 \\
& \proj ($\tau$ = 3.0) &26.24 &21.65 &25.92 &23.78 &26.28 &24.07 &25.17 \\
& \proj ($\tau$ = 15.0) &25.06 &20.67 &24.64 &22.66 &23.43 &22.51 &22.97 \\
\midrule[0.05em]
\multirow{9}{*}{SSIM $\uparrow$} & 3DGS &0.724 &0.797 &0.839 &0.818 &0.791 &0.713 &0.752 \\
& CityGaussian &0.729 &0.814 &0.835 &0.825 &0.810 &0.784 &0.797 \\
& OctreeGS &0.646 &0.771 &0.834 &0.802 &0.809 &0.720 &0.765 \\
& HierarchicalGS ($\tau$ = 0.0) &0.814 &0.675 &0.838 &0.757 &0.817 &0.757 &0.787 \\
& HierarchicalGS ($\tau$ = 3.0) &0.811 &0.674 &0.837 &0.755 &0.812 &0.754 &0.783 \\
& HierarchicalGS ($\tau$ = 15.0) &0.771 &0.639 &0.793 &0.716 &0.706 &0.666 &0.686 \\
& \proj ($\tau$ = 0.0) &0.814 &0.675 &0.838 &0.757 &0.817 &0.757 &0.787 \\
& \proj ($\tau$ = 3.0) &0.812 &0.674 &0.837 &0.755 &0.812 &0.754 &0.783 \\
& \proj ($\tau$ = 15.0) &0.772 &0.627 &0.787 &0.707 &0.689 &0.653 &0.671 \\
\midrule[0.05em]
\multirow{9}{*}{LPIPS $\downarrow$} & 3DGS &0.430 &0.264 &0.274 &0.269 &0.304 &0.356 &0.330 \\
& CityGaussian &0.344 &0.212 &0.234 &0.223 &0.233 &0.243 &0.238 \\
& OctreeGS &0.415 &0.275 &0.219 &0.247 &0.269 &0.274 &0.272 \\
& HierarchicalGS ($\tau$ = 0.0) &0.250 &0.246 &0.206 &0.226 &0.256 &0.296 &0.276 \\
& HierarchicalGS ($\tau$ = 3.0) &0.254 &0.248 &0.208 &0.228 &0.261 &0.299 &0.280 \\
& HierarchicalGS ($\tau$ = 15.0) &0.316 &0.332 &0.279 &0.306 &0.358 &0.376 &0.367 \\
& \proj ($\tau$ = 0.0) &0.250 &0.246 &0.206 &0.226 &0.256 &0.296 &0.276 \\
& \proj ($\tau$ = 3.0) &0.254 &0.248 &0.208 &0.228 &0.261 &0.299 &0.280 \\
& \proj ($\tau$ = 15.0) &0.324 &0.351 &0.311 &0.331 &0.365 &0.384 &0.375 \\
\bottomrule[0.15em]
\end{tabular}
}
\label{tab:overall_tbl1}
\end{table*}

\begin{figure*}[t]
    \centering
    \includegraphics[width=0.9\textwidth]{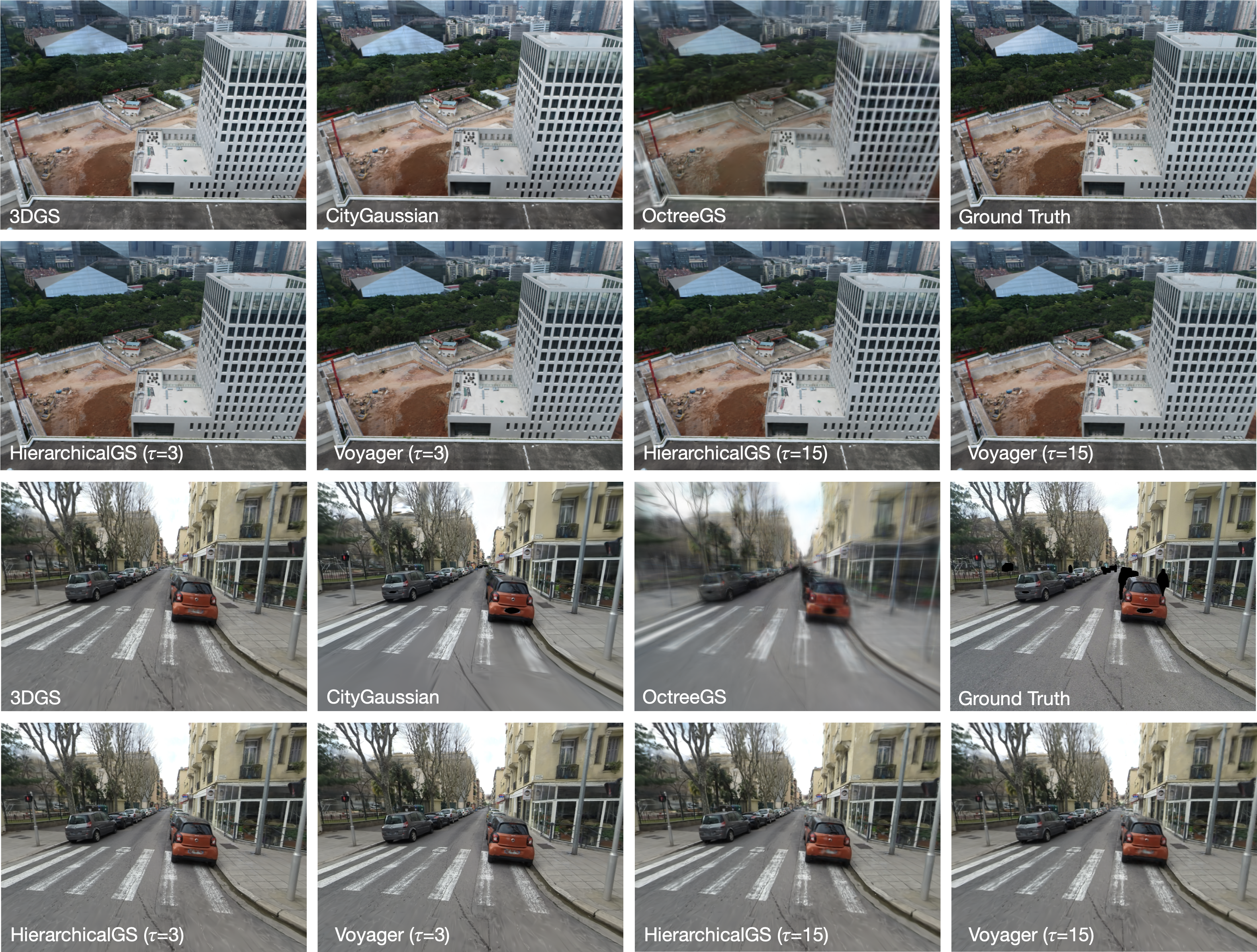}
    \caption{Qualitative comparison of \proj against other baselines~\cite{liu2024citygaussian, kerbl2024hierarchical, kerbl20233d, ren2024octree} on \textit{SciArt} and \textit{SmallCity}. \textbf{More results in the supplementary video.}}
    \label{fig:examples}
\end{figure*}

\begin{figure*}[t]
    \centering
    \includegraphics[width=0.9\textwidth]{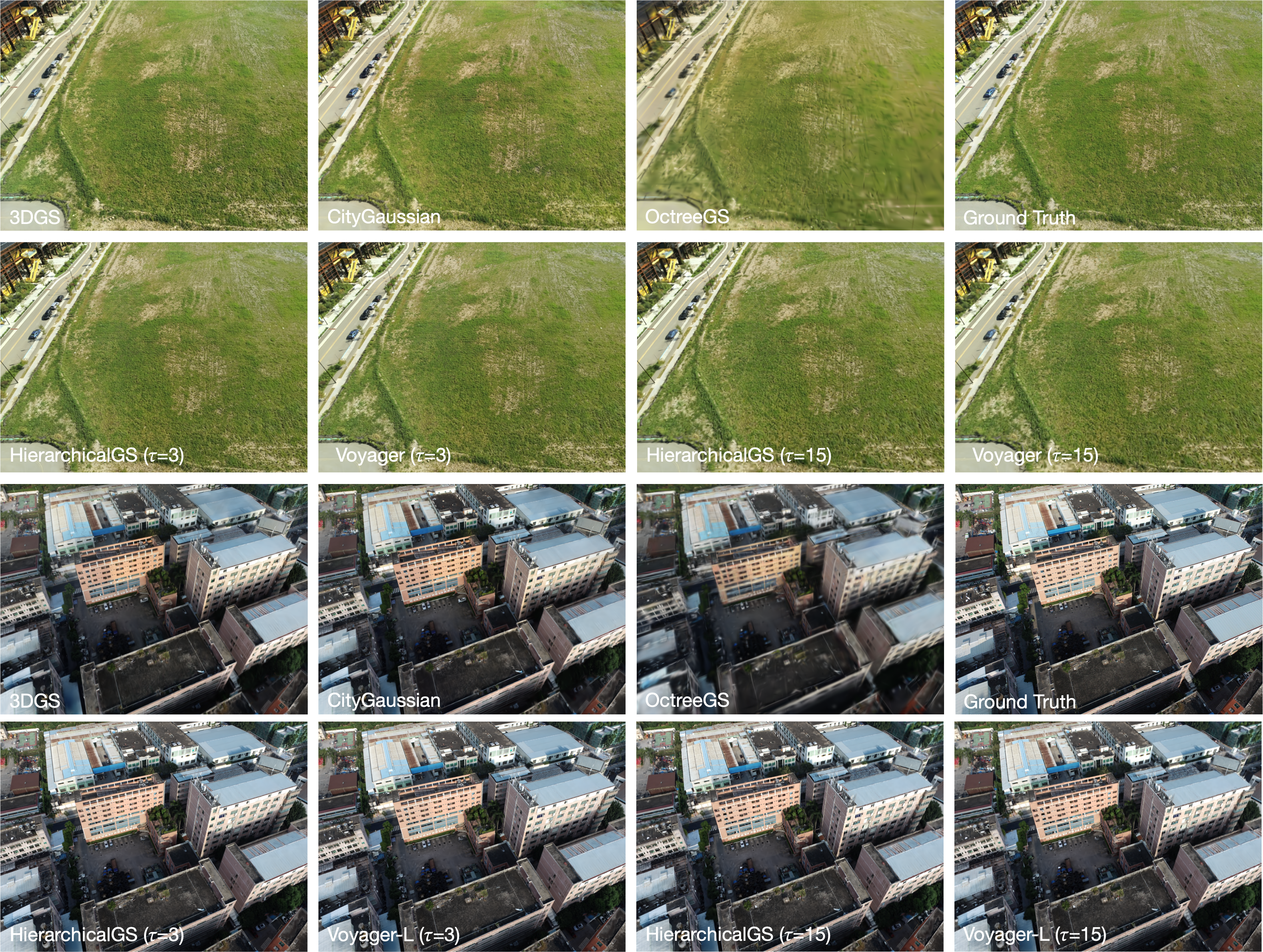}
    \caption{Qualitative comparison of \proj against other baselines~\cite{liu2024citygaussian, kerbl2024hierarchical, kerbl20233d, ren2024octree} on \textit{Building} and \textit{Residence}. \textbf{More results in the supplementary video.}}
    \label{fig:examples}
\end{figure*}

\begin{figure*}[t]
    \centering
    \includegraphics[width=0.9\textwidth]{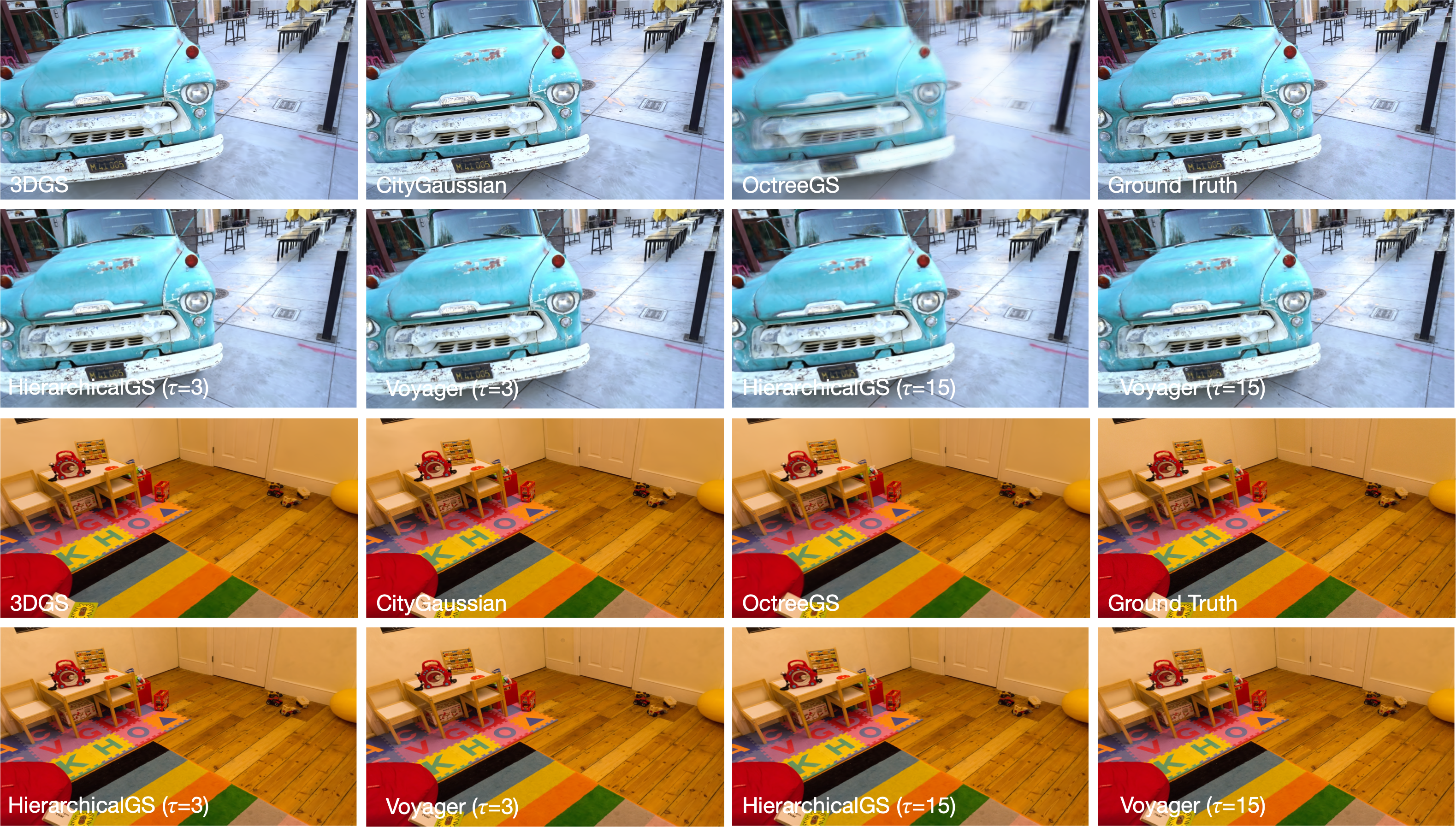}
    \caption{Qualitative comparison of \proj against other baselines~\cite{liu2024citygaussian, kerbl2024hierarchical, kerbl20233d, ren2024octree} on small-scale datasets. \textbf{More results in the supplementary video.}}
    \label{fig:examples}
\end{figure*}

\begin{figure*}[t]
    \centering
    \includegraphics[width=0.9\textwidth]{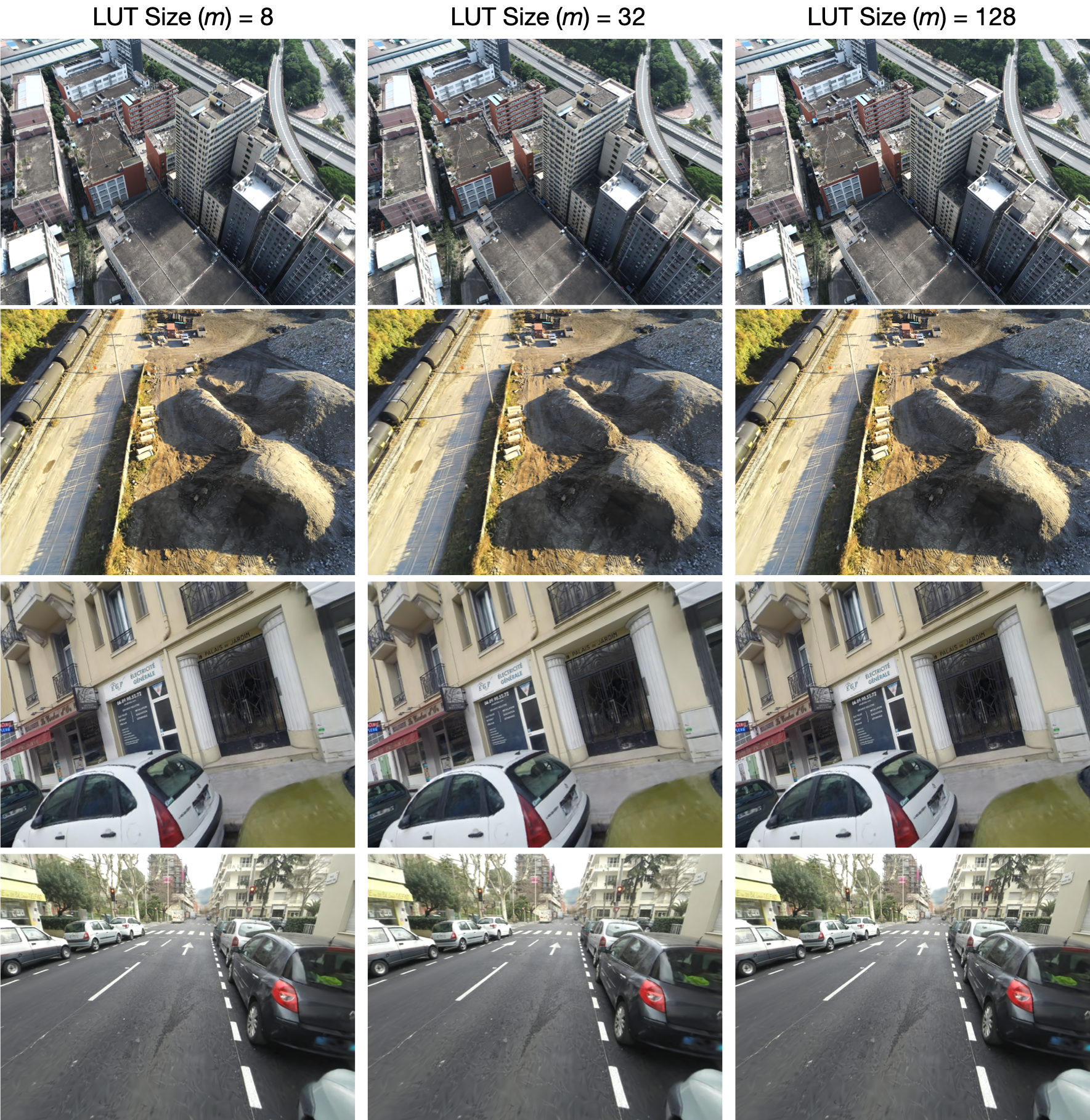}
    \caption{Qualitative results of the sensitivity of rendering quality and performance to the lookup table size ($m$ intervals) in the rasterization stage.}
    \label{fig:sens_study}
\end{figure*}

\end{alphasection}

\end{document}